\documentclass[pra,showpacs]{revtex4}

\def\pa{\partial}

\def\var{\varepsilon}

\begin{document}

\title{Quantum metastability in time-periodic potentials}

\author{Chung-Chieh Lee}
\affiliation{ Department of Electronic Engineering, Tung Nan
Institute of Technology, Shenkeng, Taipei 222, Taiwan}
\author{Choon-Lin Ho}
\affiliation{Department of Physics, Tamkang University, Tamsui
25137, Taiwan}

\date{Mar 30, 2005} 

\begin{abstract}

In this paper we investigate quantum metastability of a particle
trapped in between an infinite wall and a square barrier, with
either a time-periodically oscillating barrier (Model A) or bottom
of the well (Model B). Based on the Floquet theory, we derive in
each case an equation which determines the stability of the
metastable system. We study the influence on the stability of two
Floquet states when their Floquet energies (real part) encounter a
direct or an avoided crossing at resonance.  The effect of the
amplitude of oscillation on the nature of crossing of Floquet
energies is also discussed. It is found that by adiabatically
changing the frequency and amplitude of the oscillation field, one
can manipulate the stability of states in the well.  By means of a
discrete transform, the two models are shown to have exactly the
same Floquet energy spectrum at the same oscillating amplitude and
frequency.  The equivalence of the models is also demonstrated by
means of the principle of gauge invariance.
\end{abstract}

\maketitle

\section{Introduction}

Recent developments in powerful lasers have induced great
theoretical and experimental interests in quantum systems under
the influence of periodic driving fields. For example, an active
topic in the field of atomic physics is the dynamical
stabilization of atoms by intense high-frequency laser fields
\cite{Eberly1993,Su1996,Druten1997,Faria1999}. Scattering of
particles by a time-periodic potential is frequently used as a
model for the photon-assisted tunneling and other quantum
transport problems
\cite{Landauer1982,Wagner1994,Berman1996,Reichl1999,Henseler2001}.
The interesting phenomena of coherent destruction of tunneling in
a periodically driven two-level systems has also been discussed
extensively in the last decade \cite{Hanggi1991,Grifoni1998}.

Although diverse phenomena related to time-periodic potentials
have been widely investigated, in our opinion the interesting
issue of quantum decay through tunneling of a metastable system
trapped in a time-dependent potential should receive more
attention than it has thus far. Metastable systems which decay
through tunneling exist and are of interest in many areas of
studies, the most famous of which being the $\alpha$-decay of
atoms. In many realistic situations of decay processes, the
metastable systems have to be treated as being influenced by
time-dependent potentials.  For example, in some versions of the
inflationary models of the early universe \cite{Guth1981},
inflation is governed by a Higgs field trapped in a metastable
state. Inflation ends when the metastable state decays to the true
ground state of the universe. During inflation the universe
expands exponentially. It is thus obvious that the metastable
state of the Higgs field is trapped in a rapidly varying
potential.  Unfortunately, in these inflationary models the true
forms of the time-dependent potentials are determined by the
dynamics of the system itself. This inherent difficulty makes the
complete analysis of time-dependent decay of the Higgs field
rather difficult.  It is therefore desirable to gain some insights
first by studying metastability in various types of time-dependent
potentials in simple quantum-mechanical models.

In our previous work \cite{Lee2002} we have proposed a simple
metastable system with a moving potential which has height and
width scaled in a specific way introduced by Berry and Klein
\cite{Berry1984}. In that model we found that a small but finite
nondecay probability could persist at large time limit for an
expanding potential. Another class of time-dependent metastable
system had been discussed by Fisher\cite{Fisher1988}, in which he
considered a generic class of static metastable well subjected to
a weak time-periodic force. As the strength of the time-periodic
force was taken to be small, it could be treated as a small
time-dependent perturbation term. Fisher modified the standard
(time-independent) WKBJ approach to include a weak time-dependent
perturbation. He found that the weak time-periodic force always
enhances the time average of the decay rate of the system.  The
potential considered by Fisher has a number of oscillator-like
levels near its minimum.  The opposite situation where only two
levels are present was considered by Sokolovski \cite{Sokolovski},
who studied the effect of a small AC field mixing two levels in
the well on the tunneling rate in a semiclassical framework.

Recently, we have considered the problem of how the metastability
of a quantum system used to model the $\alpha$-decay was
influenced by an oscillating barrier \cite{Ho2004}.  The Floquet
formalism was adopted in our analysis, so that the periodic field
needs not be treated perturbatively, and the number of states in
the potential is not restricted.  Our results show that an
oscillating potential barrier generally makes a metastable system
decay faster. However, the existence of avoided crossings of
metastable states can switch a less stable state to a more stable
one, and vice versa.  It is also found that increasing the
amplitude of the oscillating field may change a direct crossing of
states into an avoided one.  Hence, by a combination of adiabatic
changes of the frequency and the amplitude of the oscillating
barrier, one can manipulate the stability of different states in a
quantum potential.  If in the static well there exists a bound
state, then it is possible to stabilize a metastable state by
adiabatically changing the oscillating frequency and amplitude of
the barrier so that the unstable state eventually crosses over to
the stable bound state.  In this connection we note here that a
related problem was studied in \cite{LN}, where the authors
consider trapping an electron in thermal equilibrium within an
infinite wall and a barrier which grows in time.  There the
confinement of particle in the well is achieved through the
increase in the barrier height, whereas in our case it is
accomplished mainly through the tuning up of the frequency of the
oscillating barrier.

In \cite{Ho2004} our main concern was the effect of avoided
crossing on the quantum metastability in a periodically
oscillating potential.  In this paper we shall present a detailed
study of the effect of direct crossing on the stability of a
metastable system, and of how the oscillating amplitude affects
the nature of direct/avoided crossings. We consider a simple
quantum system in which a particle is trapped in between an
infinite wall and a rectangular barrier. Two versions of the
system are discussed.  In Model A we consider the potential with
an oscillatory height, while in Model B we make the bottom of the
well oscillate instead. As in our previous work, the Floquet
approach is adopted here. In each model, an equation which relates
the Floquet quasienergy of the driven system to the amplitude and
frequency of the time-periodic potential is obtained.  This
equation is solved numerically and the stability of the system is
investigated as the frequency and amplitude of the oscillating
field vary.  To our surprise, the two models, very different in
nature, turn out to possess the same Floquet energy spectrum (and
thus the same stability).  A discrete transform connecting the two
models is found which explains the exactness of the Floquet
spectrum.  A physical understanding of the equivalence of the two
models is also given based on the principle of gauge invariance.

The organization of the paper is as follows. In Sect.~II, we will
first briefly describe a simple metastable model of
$\alpha$-decay. Sect.~III then introduces a modified model to
include a time-periodically oscillating barrier (Model A). We will
also describe the Floquet theory  which we adopt in our analysis.
The effects of the time-periodically oscillating barrier on the
quantum metastability of the system is studied numerically in
Sect.~IV. Model B, which has the bottom of the well oscillating,
is described in Sect.~V. Sect.~VI and Sect.~VII present,
respectively, the discrete transform and the gauge transformation
that connect the two models.  Sect.~VIII summarizes the paper.

\section{Simple static model of quantum metastable system}

A simple one-dimensional model to describe the phenomena of
$\alpha$-decay is that of a particle trapped in between an
infinite potential wall and a finite barrier \cite{Bohm1989}. The
barrier potential of such a metastable system can be expressed as
\begin{eqnarray}
 V(x)=\left\{
  \begin{array}{lll}
    \infty~, & x<0~, \\
    V_{0}>0~, & a\leq x\leq b~, \\
    0~, & 0\leq x<a~,\ \mbox{and}\ x>b~.
  \end{array} \right.
  \label{static potential}
\end{eqnarray}
The wave function of the system has the well-known form
$\Psi(x,t)=\psi(x)exp(-iEt/\hbar)$, where $\psi(x)$ satisfies the
time-independent Schr\"{o}dinger equation,
\begin{eqnarray}
 \left[-\frac{\hbar^2}{2m}\frac{d^2}{d  x^2}+V(x)\right]
 \psi(x)=E\psi(x)~.
\end{eqnarray}
The wave function can be expressed in each regions as
\begin{eqnarray}
 \psi(x) = \left\{ \begin{array}{lll}
 A\sin(kx)+B\cos(kx)~,& 0\leq x< a~,& k=\sqrt{2mE/\hbar^2}~, \\
 Ce^{qx}+De^{-qx}~,& a\leq x\leq b~,& q=
 \sqrt{2m(V_{0}-E)/\hbar^2}~, \\
 Fe^{ikx}+Ge^{-ikx}~,& b<x~.&
 \end{array}\right.
\end{eqnarray}

Besides the ordinary boundary conditions which require that the
wave function and its first derivative  be continuous at the
boundaries $x=a$ and $b$, we could also impose another two
boundary conditions on the wave function by physical
consideration. One is $B=0$ as the wave function must vanish at
the origin $(x=0)$; the other is $G=0$ since there is no extra
nonvanishing potential in the region $x>b$ to reflect wave from
the right-hand side of the barrier. This later condition was first
introduced by Gamow \cite{Gamow1928}, Curney and Condon
\cite{Curney1929}, and is usually called the ``Gamow's outgoing
wave boundary condition" in the literature. With these boundary
conditions, we can eliminate the remaining coefficients
$(A,C,D,F)$ of the wave function and obtain an equation that
determines the eigenenergy of the system,
\begin{eqnarray}
 \frac{q}{k}\tan ka +1 =\frac{q+ik}{q-ik}
 \left(\frac{q}{k}\tan ka -1\right)e^{-2q(b-a)}~.
 \label{static result}
\end{eqnarray}
By solving the above equation numerically for a given barrier
height $V_{0}$, one finds that the solution of Eq.(\ref{static
result}) gives complex eigenenergy, $E=E_{0}-i\Gamma/2$.  The
imaginary part is related to the decay rate of the state. Despite
some conceptual difficulties concerning the nature of complex
energy,  this simple approach by imposing Gamow's outgoing wave
boundary condition successfully describes the exponential decay of
metastable systems \cite{Bohm1989,Bohm1994}.

\section{Model A: The static model with a time-periodically
oscillating barrier}

We now modify the static model to include a time-periodically
oscillating barrier,
\begin{eqnarray}
 V(x,t)=\left\{
  \begin{array}{lll}
    \infty~, & x<0~, \\
    V_{0}+V_{1}\cos(\omega t)~, & a\leq x\leq b~, \\
    0~, & 0\leq x<a~,\ \mbox{and}\ x>b~,
  \end{array} \right.
  \label{potential}
\end{eqnarray}
where $V_0>0$ and $0<V_1<V_0$. Unlike the static case, in this
time-dependent model we have to solve the time-dependent
Schr\"{o}dinger equation,
\begin{eqnarray}
 i\hbar\frac{\pa \Psi(x,t)}{\pa t}
 =\left[-\frac{\hbar^2}{2m}\frac{\pa^2}{\pa x^2}+V(x,t)\right]
 \Psi(x,t)~.
 \label{Schrodinger}
\end{eqnarray}

A similar time-periodically oscillating potential (without the
infinite wall) has been used to investigate electrons scattering
in a time-dependent potential by Li and Reichl \cite{Reichl1999}.
They employed the Floquet formalism to set up the S-matrix of the
driven system. The advantage of the Floquet formalism is that one
could do away with any restriction of the amplitude and frequency
of the driving force. Here we will mainly follow the procedure
developed in \cite{Reichl1999} to construct the wave function of
our model of metastable system.

The Floquet theorem states that the wave function of a
time-periodic system has the form $\Psi_{\var}(x,t)=e^{-i\var
t/\hbar}\Phi_{\var}(x,t)$, where $\Phi_{\var}(x,t)$ is a periodic
function $\Phi_{\var}(x,t)=\Phi_{\var}(x,t+T)$ with the period
$T=2\pi/\omega$, and $\var$ is the Floquet quasienergy, or simply
the Floquet energy. It should be noted that the Floquet energy is
determined only modulo $\hbar\omega$. For if $\{
\var,\Phi_{\var}\}$ is a solution of the Schr\"{o}dinger equation,
then $\{ \var'=\var+n\hbar\omega,
\Phi_{\var'}=\Phi_{\var}\exp(in\omega t)\}$ is also a solution for
any integer $n$. But they are physically equivalent as the total
wave function $\Psi_{\var}$ is the same \cite{Sambe1973}. All
physically inequivalent states can be characterized by their
reduced Floquet energies in a zone with a width $\hbar\omega$. We
therefore consider solutions of $\var$ only in the first Floquet
zone, i.e. $\var\in [0,\hbar\omega)$.

The Schr\"{o}dinger equation inside the region of the oscillating
barrier $a\leq x\leq b$ is
\begin{eqnarray}
 -i\hbar\frac{\pa}{\pa t}\Phi_{\var}(x,t)-\frac{\hbar^2}{2m}
 \frac{\pa^2}{\pa
 x^2}\Phi_{\var}(x,t)+\left[V_{0}+V_{1}\cos (\omega t)\right]
 \Phi_{\var}(x,t)
 =\var \Phi_{\var}(x,t)~.
 \label{equiv Schr}
\end{eqnarray}
Letting $\Phi_{\var}(x,t)=g(x)f(t)$, Eq.(\ref{equiv Schr}) can be
separated into
\begin{eqnarray}
 -\frac{\hbar^2}{2m}\frac{d^2g(x)}{dx^2}+V_{0}g(x)=Eg(x)~,
 \label{x-dep}
\end{eqnarray}
and
\begin{eqnarray}
 i\hbar\frac{d f(t)}{dt}-V_{1}\cos(\omega t)f(t)=(E-\var)f(t)~,
 \label{t-dep}
\end{eqnarray}
where $E$ is a constant to be determined from the periodicity of
the function $f(t)$, i.e.  $f(t+T)=f(t)$.  Eq.(\ref{t-dep}) can be
integrated to give
\begin{eqnarray}
 f(t) = e^{-i(E-\var)t/\hbar}e^{-iV_{1}\sin(\omega t)/\hbar\omega}~,
 \label{f(t)_1}
\end{eqnarray}
where we have set the initial condition as $f(0)=1$. The second
factor in Eq.(\ref{f(t)_1}) can be rewritten using the identity
\cite{AS}
\begin{eqnarray}
 e^{-iV_{1}\sin(\omega t)/\hbar\omega}=\sum_{n=-\infty}^{\infty}J_{n}
 \left(\alpha\right)
 e^{-in\omega t}
 \ \ \ ;\ \ \
 \alpha=\frac{V_{1}}{\hbar\omega}~,\label{Bessel}
\end{eqnarray}
where $J_{n}(\alpha)$ is the Bessel function of the first kind.
Eq.(\ref{f(t)_1}) then reduces to
\begin{eqnarray}
 f(t) = e^{-i(E-\var)t/\hbar}\sum_{n=-\infty}^{\infty}J_{n}
 \left(\alpha\right)
 e^{-in\omega t}~.
 \label{f(t)}
\end{eqnarray}
The periodic condition $f(t+T)=f(t)$, which reads
\begin{eqnarray}
 f(t+T) &=& e^{-i(E-\var)(t+T)/\hbar}\sum_{n=-\infty}^{\infty}J_{n}
 \left(\alpha\right)
 e^{-in\omega (t+T)}~, \bigskip \\
 &=& e^{-i(E-\var)t/\hbar}\sum_{n=-\infty}^{\infty}J_{n}
 \left(\alpha\right)  e^{-in\omega t}e^{-i(E-\var+n\hbar\omega)
 T/\hbar}~, \bigskip \\ &=& f(t)~,
\end{eqnarray}
requires that $(E-\var+n\hbar\omega)T/\hbar=2m\pi$ ($m=$integer).
So the constant $E$ can only assume the following values
\begin{eqnarray}
 E_{l}=\var+l\hbar\omega  \ \ ,\ \  l=m-n=\mbox{integer}~.
\end{eqnarray}
The allowed values of $E$ indicate that the time-periodic
potential will generate side-band structures with
$E_{l}=\var+l\hbar\omega$ as the ``side-band energy" of the $l$-th
side-band.

Knowing the allowed values of the constant $E_{l}$, we can now
solve Eq.(\ref{x-dep}) to get
\begin{eqnarray}
 g(x)=\sum_{l}a_{l}e^{q_{l}x}+b_{l}e^{-q_{l}x}  \ \ ,\ \
 q_{l}=\sqrt{2m(V_{0}-\var-l\hbar\omega)/\hbar^2}~.
 \label{g(x)}
\end{eqnarray}
Combining Eq.(\ref{f(t)}) and Eq.(\ref{g(x)}), we obtain the wave
function inside the region of the oscillating barrier $a\leq x\leq
b$,
\begin{eqnarray}
 \Psi_{I\!I}(x,t) &=&  e^{-i\var t/\hbar}\sum_{n}\sum_{l}
 \left(a_{l}e^{q_{l}x}+b_{l}  e^{-q_{l}x}\right)
 J_{n-l}\left(\alpha\right)e^{-in\omega t}~,\nonumber
  \\ &\equiv&  \sum_{n}\sum_{l}\left(a_{l}e^{q_{l}x}+b_{l}
  e^{-q_{l}x}\right)  J_{n-l}\left(\alpha\right)
  e^{-iE_{n}t/\hbar}~.
 \label{wf-2}
\end{eqnarray}
Having gotten the wave function inside the region $a\leq x\leq b$,
the continuity of the wave function and its first derivative at
the boundaries $x=a$ and $b$ at any time requires that outside the
region $a\leq x\leq b$ there are nonvanishing wave functions on
each side-band channel. Therefore the wave function in the region
$0\leq x<a$ is
\begin{eqnarray}
 \Psi_{I}(x,t) =
 \sum_{n}A_{n}\sin(k_{n}x)e^{-iE_{n}t/\hbar}~,
 \ k_{n}=\sqrt{2mE_{n}/\hbar^2}~,
 \label{wf-1}
\end{eqnarray}
which vanishes at the origin, $\Psi_{I}(0,t)=0$. On the other
hand, the Gamow's outgoing wave boundary condition suggests that
the wave function in the region $x>b$ to have the form
\begin{eqnarray}
 \Psi_{I\!I\!I}(x,t)=
 \sum_{n}t_{n}e^{ik_{n}x}e^{-iE_{n}t/\hbar}~.
 \label{wf-3}
\end{eqnarray}

We can obtain relations among the coefficients $A_{n},a_{n},b_{n}$
and $t_{n}$ by matching the boundary conditions at $x=a$ and $b$
for each side-band. The results are
\begin{eqnarray}
 A_{n}\sin(k_{n}a) &=&
 \sum_{l}\left(a_{l}e^{q_{l}a}+b_{l}e^{-q_{l}a}\right)
 J_{n-l} (\alpha)~,  \label{b.c.1}  \\
 k_{n}A_{n}\cos(k_{n}a) &=&  \sum_{l}q_{l}\left(a_{l}e^{q_{l}a}-b_{l}
 e^{-q_{l}a}\right)J_{n-l} (\alpha)~,\label{b.c.2} \\
 t_{n}e^{ik_{n}b} &=&  \sum_{l}\left(a_{l}e^{q_{l}b}+b_{l}
 e^{-q_{l}b}\right)J_{n-l} (\alpha)~,\label{b.c.3} \\
 ik_{n}t_{n}e^{ik_{n}b} &=&  \sum_{l}q_{l}\left(a_{l}
 e^{q_{l}b}-b_{l}e^{-q_{l}b}\right)J_{n-l} (\alpha)~.
 \label{b.c.4}
\end{eqnarray}
The Floquet energy is determined from these relations by demanding
nontrivial solutions of the coefficients.  In practice, however,
we must truncate the above equations to a finite number of
side-bands, eg., $n=0, \pm1,\ldots,\pm N$ . The number $N$ is
determined by the frequency and the strength of the oscillation
according to $N>V_{1}/\hbar\omega$
\cite{Berman1996,Reichl1999,Ho2004}.

To determine the Floquet energy $\var$, we first separate the
boundary conditions for the central band $n=0$ from the other
side-bands $l\neq 0$ in Eq.(\ref{b.c.1})-(\ref{b.c.4}) (we set
$A_{0}=1$ for simplicity):
\begin{eqnarray}
 \sin(k_{0}a) &=&
 \left(a_{0}e^{q_{0}a}+b_{0}e^{-q_{0}a}\right)J_{0}(\alpha)+
 \sum_{l\neq 0}\left(a_{l}e^{q_{l}a}+b_{l}e^{-q_{l}a}\right)
 J_{-l}(\alpha)~,  \label{b.c.1-2}  \\
 k_{0}\cos(k_{0}a) &=&  q_{0}\left(a_{0}e^{q_{0}a}-b_{0}
 e^{-q_{0}a}\right)J_{0}(\alpha)+ \sum_{l\neq 0}q_{l}\left(a_{l}
 e^{q_{l}a}-b_{l}e^{-q_{l}a}\right) J_{-l}(\alpha)~, \\
 t_{0}e^{ik_{0}b} &=&  \left(a_{0}e^{q_{0}b}+b_{0}e^{-q_{0}b}
 \right)J_{0}(\alpha)+ \sum_{l\neq 0}\left(a_{l}e^{q_{l}b}+b_{l}
 e^{-q_{l}b}\right) J_{-l}(\alpha)~, \\
 ik_{0}t_{0}e^{ik_{0}b} &=&  q_{0}\left(a_{0}e^{q_{0}b}-b_{0}
 e^{-q_{0}b}\right)J_{0} (\alpha)+ \sum_{l\neq  0}q_{l}\left(a_{l}
 e^{q_{l}b}-b_{l} e^{-q_{l}b}\right)J_{-l} (\alpha)~.
 \label{b.c.4-2}
\end{eqnarray}
The coefficients $a_{l}$ and $b_{l}$ ($l\neq 0$) in
Eq.(\ref{b.c.1-2}-\ref{b.c.4-2}) can always be related to the
coefficients $a_{0}$ and $b_{0}$  as
\begin{eqnarray}
 a_{l} &=& f_{la}(k_{0},\omega,V_{1})a_{0}+f_{lb}(k_{0},\omega,
 V_{1})b_{0}~, \label{a_l} \\
 b_{l} &=& g_{la}(k_{0},\omega,V_{1})a_{0}+g_{lb}(k_{0},
 \omega,V_{1})b_{0}~, \label{b_l}
\end{eqnarray}
where $f$'s and $g$'s are functions determined as follows.
Eliminating the $A_n$'s and $t_n$'s in
Eq.(\ref{b.c.1})-(\ref{b.c.4}), we can obtain
\begin{eqnarray}
 &&A^{-}_{n,n}e^{q_{n}a}J_{0}a_{n} +A^{+}_{n,n}e^{-q_{n}a}
 J_{0}b_{n} +\sum_{l\neq n,0}A^{-}_{n,l}e^{q_{l}a}J_{n-l}a_{l}
 +\sum_{l\neq n,0}A^{+}_{n,l}e^{-q_{l}a}J_{n-l}b_{l}
 \nonumber \bigskip \\
 &=&
 -A^{-}_{n,0}e^{q_{0}a}J_{n}a_{0}-A^{+}_{n,0}
 e^{-q_{0}a}J_{n}b_{0}~,
 \label{b.c.neq0_1}
\end{eqnarray}
and
\begin{eqnarray}
 &&B^{+}_{n,n}e^{q_{n}b}J_{0}a_{n}+B^{-}_{n,n}
 e^{-q_{n}b}J_{0}b_{n}
 +\sum_{l\neq n,0}B^{+}_{n,l}e^{q_{l}b}J_{n-l}a_{l}
 +\sum_{l\neq n,0}B^{-}_{n,l}e^{-q_{l}b}J_{n-l}b_{l}
 \nonumber \bigskip \\
 &=& -B^{+}_{n,0}e^{q_{0}b}J_{n}a_{0}-B^{-}_{n,0}
 e^{-q_{0}b}J_{n}b_{0}~,
 \label{b.c.neq0_2}
\end{eqnarray}
where
\begin{eqnarray}
 A^{\pm}_{n,l} \equiv \cos k_{n}a \pm \frac{q_{l}}{k_{n}}\sin
 k_{n}a~,~~{\rm  and}~~
 B^{\pm}_{n,l} \equiv 1 \pm i\frac{q_{l}}{k_{n}}~.\label{AB}
\end{eqnarray}
As mentioned before, in practice a truncated version of Eqs.
(\ref{b.c.neq0_1}) and (\ref{b.c.neq0_2}) has to be used.  In a
$m$ side-bands approximation, the index $n$ ranges from $-m$ to
$m$ except $0$.  Then Eqs. (\ref{b.c.neq0_1}) and
(\ref{b.c.neq0_2}) together represent a set of  $2m$ equations
with $2m$ unknowns $a_{n}$ and $b_{n}$. The solutions of these
unknowns can then be solved by the Cramer's rule \cite{Boas} in
the forms of Eq.(\ref{a_l}) and Eq.(\ref{b_l}).

Replacing the $a_l$'s and $b_l$'s by means of Eqs.(\ref{a_l}) and
(\ref{b_l}), one can rewrite Eqs.(\ref{b.c.1-2}) to
(\ref{b.c.4-2}) as
\begin{eqnarray}
 \sin(k_{0}a) &=&
 F_{1}(k_{0};\omega,V_{1})e^{q_{0}a}a_{0}+F_{2}(k_{0};\omega,V_{1})
 e^{-q_{0}a}b_{0}~, \label{b.c.1-3} \\
 k_{0}\cos(k_{0}a) &=& F_{3}(k_{0};\omega,V_{1})q_{0}e^{q_{0}a}
 a_{0}-F_{4}(k_{0}; \omega,V_{1})q_{0}e^{-q_{0}a}b_{0}~, \\
 t_{0}e^{ik_{0}b} &=& F_{5}(k_{0};\omega,V_{1})e^{q_{0}b}
 a_{0}+F_{6}(k_{0}; \omega,V_{1})e^{-q_{0}b}b_{0}~, \\
 ik_{0}t_{0}e^{ik_{0}b} &=& F_{7}(k_{0};\omega,V_{1})q_{0}
 e^{q_{0}b}a_{0}-F_{8}(k_{0}; \omega,V_{1})q_{0}e^{-q_{0}b}b_{0}~,
 \label{b.c.4-3}
\end{eqnarray}
where the coefficient $F_{i}(k_{0};\omega,V_{1})\ (i=1,\dots,8)$ are
\begin{eqnarray}
 F_{1}(k_{0};\omega,V_{1}) &=&
 J_{0}(\alpha)+e^{-q_{0}a}\sum_{l\neq 0}
 \left(f_{la}e^{q_{l}a}+g_{la}e^{-q_{l}a}\right)J_{-l}(\alpha)~,
 \label{F_1} \\
 F_{2}(k_{0};\omega,V_{1}) &=&
 J_{0}(\alpha)+e^{q_{0}a}\sum_{l\neq 0}
 \left(f_{lb}e^{q_{l}a}+g_{lb}e^{-q_{l}a}\right)J_{-l}(\alpha)~, \\
 F_{3}(k_{0};\omega,V_{1}) &=&
 J_{0}(\alpha)+e^{-q_{0}a}\sum_{l\neq 0}
 \frac{q_{l}}{q_{0}}\left(f_{la}e^{q_{l}a}-g_{la}e^{-q_{l}a}\right)
 J_{-l}(\alpha)~, \\
 F_{4}(k_{0};\omega,V_{1}) &=&
 J_{0}(\alpha)-e^{q_{0}a}\sum_{l\neq 0}
 \frac{q_{l}}{q_{0}}\left(f_{lb}e^{q_{l}a}-g_{lb}e^{-q_{l}a}\right)
 J_{-l}(\alpha)~, \\
 F_{5}(k_{0};\omega,V_{1}) &=&
 J_{0}(\alpha)+e^{-q_{0}b}\sum_{l\neq 0}
 \left(f_{la}e^{q_{l}b}+g_{la}e^{-q_{l}b}\right)J_{-l}(\alpha)~, \\
 F_{6}(k_{0};\omega,V_{1}) &=&
 J_{0}(\alpha)+e^{q_{0}b}\sum_{l\neq 0}
 \left(f_{lb}e^{q_{l}b}+g_{lb}e^{-q_{l}b}\right)J_{-l}(\alpha)~, \\
 F_{7}(k_{0};\omega,V_{1}) &=&
 J_{0}(\alpha)+e^{-q_{0}b}\sum_{l\neq 0}
 \frac{q_{l}}{q_{0}}\left(f_{la}e^{q_{l}b}-g_{la}e^{-q_{l}b}\right)
 J_{-l}(\alpha)~, \\
 F_{8}(k_{0};\omega,V_{1}) &=&
 J_{0}(\alpha)-e^{q_{0}b}\sum_{l\neq 0}
 \frac{q_{l}}{q_{0}}\left(f_{lb}e^{q_{l}b}-g_{lb}e^{-q_{l}b}\right)
 J_{-l}(\alpha)~. \label{F_8}
\end{eqnarray}
Then by eliminating the coefficients $a_{0}$, $b_{0}$ and $t_{0}$
in Eqs.(\ref{b.c.1-3}) to (\ref{b.c.4-3}), we arrive at an
equation which is a function of $\var\ (=\hbar^2k_{0}^2/2m)$,
\begin{eqnarray}
 F_{4}\frac{q_{0}}{k_{0}}\tan k_{0}a +F_{2}
 =\frac{F_{8}q_{0}+iF_{6}k_{0}}{F_{7}q_{0}-iF_{5}k_{0}}
 \left(F_{3}\frac{q_{0}}{k_{0}}\tan k_{0}a -F_{1}\right)
 e^{-2q_{0}(b-a)}~.
 \label{solution}
\end{eqnarray}
Its solution determines the Floquet energy of the metastable
system.  Comparing this equation with the corresponding one for
the static case, i.e. Eq.(\ref{static result}), we  see the effect
of the time-periodic potential is contained entirely in the
coefficients $F_{i}$.

In the next section we will solve Eq.(\ref{solution}) numerically.
Once the Floquet energy $\var$ of the metastable system is
determined, the probability of the particle still being trapped by
the potential barrier at time $t>0$, i.e.  the non-decay
probability $P(t)$, can be obtained as
\begin{eqnarray}
 P(t) &=& \frac{\int_{0}^{b}\left|\Psi(x,t)\right|^2dx}
 {\int_{0}^{b}\left|\Psi(x,0)\right|^2dx}
 \nonumber \\
 &=& e^{2Im(\var)t/\hbar} \frac{\int_{0}^{b}\left|\Phi_\var(x,t)
 \right|^2dx} {\int_{0}^{b}\left|\Phi_\var(x,0)\right|^2dx}
 \label{P(t)}\\
 &\equiv & e^{2Im(\var)t/\hbar} h(t)~,\nonumber
\end{eqnarray}
with $P(0)=1$. The imaginary part of the Floquet energy, which
enters $P(t)$ via the factor $\exp(2Im(\var)t/\hbar)$, gives a
measure of the stability of the system.  Unlike the static case,
however, here $P(t)$ is not a monotonic function of time, owing to
the time-dependent function $h(t)$. But since $h(t)$ is only a
periodic function oscillating between two values which are of
order one, the essential behavior of $P(t)$ at large times is
still mainly governed by the exponential factor.  Hence, as a
useful measure of the nondecay rate of the particle in the well,
we can use a coarse-grained nondecay probability $\bar{P}(t)$
defined as \cite{Ho2004}
\begin{eqnarray}
\bar{P}(t)\equiv  e^{2Im(\var)t/\hbar} \langle h(t) \rangle~,
\label{bP(t)}
\end{eqnarray}
where $\langle h(t) \rangle$ is the time average of $h(t)$ over
one period of oscillation.

Owing to the complexity of the coefficients
$F_{i}(k_{0};\omega,V_{1})$, the solution of Eq.(\ref{solution})
has to be obtained  numerically.  We shall present the numerical
analysis in the next section.  Here let us consider the limiting
cases of our model. From the expressions of the coefficients
$F_{i}(k_{0};\omega,V_{1})$ (Eq.(\ref{F_1}) to Eq.(\ref{F_8})), we
can easily check the coefficients $F_{i}(k_{0};\omega,V_{1})$ all
approach one when the parameter
$\alpha=V_{1}/\hbar\omega\rightarrow 0$,
\begin{eqnarray}
 \lim_{V_{1}/\hbar\omega\rightarrow 0}F_{i}(k_{0};\omega,V_{1})
 \longrightarrow 1~, \ i=1,\dots,8.
\end{eqnarray}
Hence in the limit $V_{1}\rightarrow 0$ or $\omega\rightarrow
\infty$, Eq.(\ref{solution}) is identical to Eq.(\ref{static
result}) in the static case, and the Floquet energy in these
limits are just the complex eigenenergy of the static case. This
is understandable, since in the limit $V_{1}\rightarrow 0$ the
potential becomes static. And at very high frequencies, the
dynamical time scale of the particle in the well is much larger
than the time scale of the oscillating barrier $\omega^{-1}$,
hence the height of the time-periodically oscillating barrier seen
by the particle should be its time-averaged value, namely, $<
V_{0}+V_{1}\cos \omega t
>=V_{0}$, which is independent of the time-oscillating part of the
barrier \cite{Landauer1982}.

\section{Numerical analysis of Model A}

In this section, we will investigate the quantum metastability of
Model A by finding the solutions of Eq.(\ref{solution})
numerically. We are interested in the metastability of the system
under different amplitudes and frequencies of the oscillating
barrier, since these are the experimentally controllable
parameters.  For definiteness, we take the barrier to have height
$V_{0}=10.0$ in the atomic units (a.u.) $(e=m_{e}=\hbar=1)$, and
 boundaries at $a=1$ and $b=2$ (a.u.). With these
parameters,  there is only one solution with $Re(E)<V_0$ in the
static case.  This is a metastable state having complex energy
$E_{0}/V_{0}=0.322052- 0.000110412 i$. Other solutions have
$Re(E)>V_0$ and are more unstable. We shall be interested in this
paper only in the coupling between the metastable state $E_0$ and
the next higher one with $E_1/V_0=1.11205- 0.025062i$. Coupling
between other metastable states can be considered similarly.  For
the oscillating potential, we shall solve Eq.(\ref{solution})
numerically for cases with different oscillating amplitude
$V_1/V_0$ and frequency $\omega/V_0$.

\subsection{Comparison of 2 and 3 side-bands approximations}

In this subsection we first study how good a 2 side-bands
approximation is by comparing it with the 3 side-bands
approximation.  Floquet energy as a function of the amplitude of
the oscillating barrier $V_{1}$ evaluated based on  2 side-bands
(solid line) and 3 side-bands (dotted line) approximations are
shown in Fig.(1) and Fig.(2) for $\omega/V_0=0.01$ and $0.02$,
respectively.

As mentioned before, if $\{ \var,\Phi_{\var}\}$ is a solution of
the Schr\"{o}dinger equation, then $\{ \var'=\var+n\hbar\omega,
\Phi_{\var'}=\Phi_{\var}exp(in\omega t)\}$ is also a solution for
any integer $n$. And they are physically equivalent as the total
wave function $\Psi_{\var}$ is the same.  For metastable system,
this means that all Floquet states with real parts differing by
$n\hbar\omega$ ($n=0,\pm 1,\ldots$) will have the same imaginary
part. We have checked this for $n=0,\pm1$ in the 2 and 3
side-bands approximations.

Part(a) [part(b)] of Fig.~1 and 2 show the relation of the real
(imaginary) part of the Floquet energy and the oscillating
amplitude parameter $V_{1}/V_{0}$ at fixed $\omega$. Fig.~(1-a)
and (2-a) show that if $Re(\var/V_{0})=Re(E_{0}/V_{0})=0.322052$
is the Floquet energy of the system, so are the $Re(\var/V_{0})\pm
\omega/V_{0}$ for all values of $V_1$ less than certain limit.
Fig.~(1-b) and (2-b) indicate that the imaginary part of these
solutions merge into one curve for $V_{1}/V_{0}\leq 0.01$ in
Fig.(1-b) and $V_{1}/V_{0}\leq 0.02$ in Fig.(2-b).  The divergence
of the three curves at higher $V_1$ is caused by the truncation of
the number of side-bands in numerical analysis. Fig.(1-b) shows
that the 2 side-bands approximation is good for $V_{1}/V_{0}\leq
0.01$ at $\omega/V_0=0.01$ (i.e. $\alpha\equiv V_{1}/\omega\leq
1$), and the 3 side-bands approximation is accurate for
$V_{1}/V_{0}\leq 0.02$  at $\omega/V_0=0.02$ (i.e. $\alpha\equiv
V_{1}/\omega\leq 2$). These results are consistent with the usual
requirement for the N side-bands approximation mentioned
previously, namely, $\alpha\equiv V_{1}/\omega\leq N$
\cite{Reichl1999,Berman1996,Ho2004}. Similar argument for the
accuracy in the approximation can also be checked in Fig.(2-b).
Stability of the system is determined by the imaginary part of the
Floquet energy. From part (b) of Fig.~1 and 2, we see that the
imaginary part of Floquet energy increases as $V_1$ increases.
This means that, at fixed frequency, a larger oscillating
amplitude will cause the system to be less stable.  On the other
hand, once $V_{1}$ is reduced to zero the quantum metastability of
the system will approach to the static case $(Log
(-Im(\var/V_{0})) = -3.95698)$, because, as we explained at the
end of the last section, in the limit $V_{1}\to 0$ the driven
system reduces to the static case.

\subsection{Floquet energy as function of the frequency
and amplitude of the oscillating field}

In this subsection, we will investigate the effects of the
oscillating frequency and amplitude on the quantum metastability
of the system in the 2 side-bands approximation. As we have
learned in the previous subsection, the 2 side-bands approximation
is good enough for $\omega
>V_1\ (\alpha^{-1}>1)$.

Fig.~(3-a) and (3-b) present, respectively, the relation of the
real and imaginary parts of the Floquet energy and the driving
frequency with the parameters $V_{0}=10.0$, $a=1,b=2$, and
$V_{1}=1.0$ in the atomic units. The solutions of
Eq.(\ref{solution}) have the form $\var=\var_0 + n\omega$
($n=0,\pm 1, \pm 2,\ldots$), with $Re(\var_0)$ (the horizontal
branch) lie close to the energies $Re(E_0)$ and $Re(E_1)$ in the
static potential. That is, these branches of $Re(\var)$ emanate
from either $Re(E_0)$ or $Re (E_1)$ at $\omega=0$. Branches
emerging from the same point have the same imaginary part.
Numerical results show that, with the barrier oscillating, these
states become less stable.  For simplicity, in Fig.~(3a) we show
only two branches emerging from the two states with $Re(E_0)$
(solid curves for $n=0,1$) and $Re (E_1)$ (dotted curves for
$n=0,-1$) in the static case. As mentioned before,  solutions can
always be reduced to the first Floquet zone, $Re(\var)$ (modulo
$\omega$), which are points under the line $Re(\var)=\omega$.

Figure (3-b) shows that the imaginary parts the Floquet energies
of the two states fluctuate slightly as the frequency changes.  At
frequency close to the resonance frequency $\omega\approx
\omega_R\equiv  Re(E_1)-Re(E_0)\approx 0.79 V_0$, a direct
crossing of the real parts of the Floquet energies of the two
states occurs. This is indicated by the point $a$ in the first
Floquet zone, or equivalently, point $b$ in the second Floquet
zone.  That it is a direct crossing is shown more clearly with a
refined scale in Fig.~(3-c) and (3-d). We see that for the first
metastable state (solid curve), $Re(\var)$ exhibits around
$\omega_R$ a ``Fano" resonance pattern, where a sharp dip is
followed by a peak.  For the second metastable state (dotted
curve), an ``anti-Fano" pattern was seen instead.  In the
neighborhood of $\omega_R$, the first metastable state becomes
less unstable (point $c$), while the second state appears to be
slightly more stable (point $d$). Beyond $\omega_R$, the stability
of the two states revert to their respective orders of magnitude.
Similar behavior is also observed at the point of direct crossing
in the system discussed in \cite{Ho2004}.

Figures 4 to 7 present similar graphs as in Fig.~3, but with
$V_1/V_0=1.6,1.7,2.0$ and $4.0$, respectively (other parameters
being the same).  In Fig.~4 we see that, with $V_1/V_0$ increases
up to $1.6$,  the system still behaves essentially in the same way
as when $V_0/V_1=1.0$. Only the first metastable state becomes
less stable, while the second metastable state tends to be more
stable at the point of direct crossing.

The behavior of the system changes qualitatively, when $V_0/V_1$
increases beyond a critical point.  At $V_1/V_0=1.7$, the crossing
between the two states changes from a direct crossing into an
avoided crossing, as depicted in Fig.~5. Increasing $V_1/V_0$
further only enhances the difference of the Floquet energies at
the point of avoided crossing (Fig.~6 and 7).  At and beyond the
avoided crossing, the two states exchange stability: the second
and originally less stable state becomes more stable than the
first state, which now assumes a stability which is of similar
order of magnitude as that of the second state before the
crossing.  Such phenomenon has been previously discussed in
\cite{Ho2004}.

The above results indicate the possibility of controlling
stability of states in a quantum well.  One may tune up the
frequency adiabatically up to the point where the real parts of
the Floquet energy of the two states meet.  If the crossing is a
direct crossing, then one can increase the magnitude of the
oscillation $V_1$ of the barrier before the crossing until the
direct crossing turns into an avoided one. When
 the frequency is increased further to pass beyond the avoided
crossing, the two states exchange stability. Then by reducing
$V_1$ to zero, the two states in the static well are interchanged.
If the lowest energy in the static well is a stable bound state
(as in the case discussed in \cite{Ho2004}), the above procedure
can turn an upper metastable state into a stable one.

\section{Model B: The static model with a time-periodically
oscillating bottom}

Let us now consider a variant of the above model, namely, the
static model with a time-periodically oscillating bottom instead
of an oscillating barrier.  The potential can be expressed as
\begin{eqnarray}
 V(x,t)=\left\{
  \begin{array}{lll}
    \infty~, & x<0~, \\
    V_{1}\cos(\omega t)~, & 0\leq x<a~, \\
    V_{0}~, & a\leq x\leq b~, \\
    0~, & x>b~.
  \end{array} \right.
\end{eqnarray}
The wave function inside the barrier well, i.e., $0\leq x\leq a$,
can be obtained by the similar procedure as in Sect.~III, except
that we have to impose the boundary condition $\Psi'_{I}(x=0,t)=0$
in this case. The result is
\begin{eqnarray}
 \Psi'_{I}(x,t) &=&
 \sum_{l}\sum_{n}A'_{n}\sin(k_{n}x)J_{l-n}(\alpha)
 e^{-i(\var+l\hbar\omega)t/\hbar}~, \nonumber \bigskip \\
 &=&
 \sum_{l}\sum_{n}A'_{n}\sin(k_{n}x)J_{l-n}(\alpha)
 e^{-iE_{l}t/\hbar}~, \label{wf-B1}
\end{eqnarray}
where $k_{n}=\sqrt{2m(\var+n\hbar\omega)}/\hbar$. Continuity of
the wave function at the boundaries $x=a$ and $b$ requires that
the wave function  has the form
\begin{eqnarray}
 \Psi'_{I\!I}(x,t) &=&
 \sum_{l}\left(a'_{l}e^{q_{l}x}+b'_{l}e^{-q_{l}x}\right)
 e^{-iE_{l}t/\hbar}  \label{wf-B2}
\end{eqnarray}
in the region $a\leq x<b$, where
$q_{l}=\sqrt{2m(V_0-\var-l\hbar\omega)}/\hbar$, and
\begin{eqnarray}
 \Psi'_{I\!I\!I}(x,t)=
 \sum_{l}t'_{l}e^{ik_{l}x}e^{-iE_{l}t/\hbar}~,
  \label{wf-B3}
\end{eqnarray}
in the region $x\geq b$, where the Gamow's outgoing wave boundary
condition has been imposed.

The relations among the coefficients $A'_{n}, a'_{n}, b'_{n}$ and
$t'_{n}$ can be obtained by matching the boundary conditions at
$x=a$ and $b$.  This leads to
\begin{eqnarray}
 \sum_{n}A'_{n}\sin(k_{n}a)J_{l-n}(\alpha) &=&
 a'_{l}e^{q_{l}a}+b'_{l}e^{-q_{l}a}~,
 \label{b.c.B-1} \\
 \sum_{n}k_{l}A'_{n}\cos(k_{n}a)J_{l-n}(\alpha) &=&
 q_{l}\left(a'_{l}e^{q_{l}a}-b'_{l}e^{-q_{l}a}\right)~,
 \label{b.c.B-2} \\
 t'_{l}e^{ik_{l}b} &=&
 a'_{l}e^{q_{l}b}+b'_{l}e^{-q_{l}b}~,
 \label{b.c.B-3} \\
 ik_{l}t'_{l}e^{ik_{l}b} &=&
 q_{l}\left(a'_{l}e^{q_{l}b}-b'_{l}e^{-q_{l}b}\right)~,
 \label{b.c.B-4}
\end{eqnarray}
Eliminating $t'_{l}$ from Eq.(\ref{b.c.B-3}) and
Eq.(\ref{b.c.B-4}) gives the relation between $a'_{l}$ and
$b'_{l}$,
\begin{eqnarray}
 b'_{l}=-\frac{1+iq_{l}/k_{l}}{1-iq_{l}/k_{l}}e^{2q_{l}b}a'_{l}~.
\end{eqnarray}
Using this relation to replace $b'_{l}$ in Eq.(\ref{b.c.B-1}) and
Eq.(\ref{b.c.B-2}), we obtain
\begin{eqnarray}
 \sum_{n}A'_{n}\sin(k_{n}a)J_{l-n}(\alpha) =
 e^{q_{l}b}\left(
 e^{-q_{l}(b-a)}-\frac{1+iq_{l}/k_{l}}{1-iq_{l}/k_{l}}e^{q_{l}(b-a)}
 \right)a'_{l}~,
\end{eqnarray}
and
\begin{eqnarray}
 \sum_{n}A'_{n}k_{n}\cos(k_{n}a)J_{l-n}(\alpha) =
 q_{l}e^{q_{l}b}\left(
 e^{-q_{l}(b-a)}+\frac{1+iq_{l}/k_{l}}{1-iq_{l}/k_{l}}e^{q_{l}(b-a)}
 \right)a'_{l}~,
\end{eqnarray}
respectively. These two equations can be combined by eliminating
the coefficients $a'_{l}$, giving
\begin{eqnarray}
 \sum_{n}\left[
 A^{-}_{n,l} e^{-q_{l}(b-a)}
 -\frac{B^{+}_{l,l}}{B^{-}_{l,l}}A^{+}_{n,l}e^{q_{l}(b-a)}
 \right]k_{n}J_{l-n}(\alpha)A'_{n}=0~,
 \label{b.c.B}
\end{eqnarray}
where the notation $A^{\pm}_{n,l}$ and $B^{\pm}_{n,l}$ are as
defined in Model A, Eq.~(\ref{AB}).

We now show how to determine the Floquet energy of the metastable
system from Eq.(\ref{b.c.B}).  For a general amplitude $V_{1}\neq
0$, Eq.(\ref{b.c.B}) in the m side-bands approximation is
\begin{eqnarray}
 \sum_{n=-m}^{m}\left[
 A^{-}_{n,l} e^{-q_{l}(b-a)}
 -\frac{B^{+}_{l,l}}{B^{-}_{l,l}}A^{+}_{n,l}e^{q_{l}(b-a)}
 \right]k_{n}J_{l-n}(\alpha)A'_{n}=0~.
 \label{b.c.BB}
\end{eqnarray}
Moving the term with the coefficient $A'_{0}$ of the above
equation to the right-hand side, we get
\begin{eqnarray}
 &&\sum_{\stackrel{n=-m}{n\neq 0}}^{m}
 \left[
 A^{-}_{n,l} e^{-q_{l}(b-a)}
 -\frac{B^{+}_{l,l}}{B^{-}_{l,l}}A^{+}_{n,l}e^{q_{l}(b-a)}
 \right]k_{n}J_{l-n}(\alpha)A'_{n}
 \nonumber \bigskip \\
 &=& -\left[
 A^{-}_{0,l} e^{-q_{l}(b-a)}
 -\frac{B^{+}_{l,l}}{B^{-}_{l,l}}A^{+}_{0,l}e^{q_{l}(b-a)}
 \right]k_{0}J_{l}(\alpha)A'_{0}~,
\end{eqnarray}
The left-hand side of the equation contains $2m$ unknown
coefficients $A'_{n\neq 0}$.  As the total number of side-band is
also $2m$ ($l=\pm 1,\cdots,\pm m$), it is sufficient to use the
Cramer's rule to obtain the relation
\begin{eqnarray}
 A'_{n\neq 0}=C_{n}(\var,V_{1},\omega)A'_{0}
\end{eqnarray}
with $C_{0}=1$. We now insert this representation of $A'_{n\neq
0}$ into Eq.(\ref{b.c.BB}). Eq.(\ref{b.c.BB}) for  the central
band ($l=0$) becomes
\begin{eqnarray}
 \left\{ \sum_{n=-m}^{m}\left[
 A^{-}_{n,0} e^{-q_{0}(b-a)}
 -\frac{B^{+}_{0,0}}{B^{-}_{0,0}}A^{+}_{n,0}e^{q_{0}(b-a)}
 \right]k_{n}J_{-n}(\alpha)C_{n}
 \right\}A'_{0}=0~.
\end{eqnarray}
Nontrivial solution of $A'_{0}$, i.e. $A'_{0}\neq 0$, requires
that
\begin{eqnarray}
 \sum_{n=-m}^{m}\left[
 A^{-}_{n,0} e^{-q_{0}(b-a)}
 -\frac{B^{+}_{0,0}}{B^{-}_{0,0}}A^{+}_{n,0}e^{q_{0}(b-a)}
 \right]k_{n}J_{-n}(\alpha)C_{n}=0~.
 \label{Final}
\end{eqnarray}
The Floquet energy of the metastable system $\var$ can then be
determined by finding the solutions of this equation.

Let us mention briefly the limiting case in which $V_{1}\to 0$
(i.e., $\alpha\to 0$). Since $J_{0}(\alpha=0)=1$ and $J_{l\neq
0}(\alpha=0)=0$, Eq.(\ref{Final}) reduces to
\begin{eqnarray}
 A^{-}_{0,0} e^{-q_{0}(b-a)}
 -\frac{B^{+}_{0,0}}{B^{-}_{0,0}}A^{+}_{0,0}e^{q_{0}(b-a)}
 =0~,
\end{eqnarray}
which, as we expect, is just the result of the static case
\begin{eqnarray}
 \frac{q_{0}+ik_{0}}{q_{0}-ik_{0}}
 \left(\frac{q_{0}}{k_{0}}\tan k_{0}a -1 \right)e^{-2q_{0}(b-a)}
 =\frac{q_{0}}{k_{0}}\tan k_{0}a +1~.
\end{eqnarray}

We had  solved Eq.(\ref{Final}) numerically for the Floquet energy
of the system.  To our surprise, the functional dependence of the
Floqeut spectrum on the amplitude and the frequency of the
oscillation turns out to be exactly the same as that of Model A.
Since the sets of equations of boundary conditions determining the
Floquet energy are so different in form for the two models , that
they should  give exactly the same solutions can be of no
coincidence. This prompted us to search for the underlying
connection between the two models.   After some efforts we
eventually realize that these models are related by a discrete
transform in the sense that Eqs.~(\ref{b.c.1})-(\ref{b.c.4}) can
be transformed into Eqs.~(\ref{b.c.B-1})-(\ref{b.c.B-4}), and vice
versa, by the discrete transform to be described below.  Since
Eq.~(\ref{b.c.2}) [(\ref{b.c.B-2})] and Eq.~(\ref{b.c.4})
[(\ref{b.c.B-4})] are obtainable from Eq.~(\ref{b.c.1})
[(\ref{b.c.B-1})] and Eq.~(\ref{b.c.3}) [(\ref{b.c.B-3})] by
differentiation with respect to the parameters $a$ and $b$,
respectively, it suffices to show the transformations between
Eqs.~(\ref{b.c.1}) and (\ref{b.c.B-1}), and between
Eqs.~(\ref{b.c.3}) and (\ref{b.c.B-3}).  This will be discussed in
the next section.  More recently, we have also realized that the
underlying physics responsible for the equivalence of the two
models is the gauge equivalence of the models, which we shall
present in Sect~VII.

\section{Discrete transform}

The aforementioned discrete transform is defined as follows.
Consider an infinite discrete sequence
$\{g_n\}=\{\ldots,g_{-1},g_0,g_1,\ldots \}$ of complex numbers
$g_n$ ($n=-\infty,\ldots,\infty$) with finite norms. Let us form
another sequence $\{g^\prime_n\}$ by the following transform based
on the Bessel functions $J_k(\alpha)$ with integral order $k$ and
argument $\alpha$:
\begin{eqnarray}
g^\prime_l \left(\alpha\right)&:=& {\cal
H}_l\left[\{g_n\},\alpha\right]\nonumber\\ &\equiv
&\sum_{n=-\infty}^{\infty} (-1)^n g_n J_{l-n}\left(\alpha\right)~,
~~~n=0,\pm 1,\pm 2,\ldots. \label{g'}
\end{eqnarray}
This transform has the interesting property that its inverse
transform is given by the same transform, i.e.  the sequence
$\{g_n\}$ is obtainable from $\{g^\prime_n\}$ by the same
transformation:
\begin{eqnarray}
g_n &=& {\cal H}_n\left[\{g^\prime_l\},\alpha\right]\nonumber\\
&=&\sum_{l=-\infty}^{\infty} (-1)^l g^\prime_l
J_{n-l}\left(\alpha\right)~. \label{g}
\end{eqnarray}

To prove this, let us substitute Eq.(\ref{g'}) into the right-hand
side of  Eq.~(\ref{g}). This gives
\begin{eqnarray}
&&\sum_{l=-\infty}^{\infty} (-1)^l g^\prime_l
J_{n-l}\left(\alpha\right) \nonumber\\
&=&\sum_{l=-\infty}^{\infty} (-1)^l
\left[\sum_{m=-\infty}^{\infty} (-1)^m g_m
J_{l-m}\left(\alpha\right)\right]
J_{n-l}\left(\alpha\right)\nonumber\\ &=&\sum_{m=-\infty}^{\infty}
(-1)^m g_m \sum_{l=-\infty}^{\infty} (-1)^l
J_{l-m}\left(\alpha\right) J_{n-l}\left(\alpha\right)\nonumber\\
&=&\sum_{m=-\infty}^{\infty} (-1)^{m+n} g_m
\sum_{l=-\infty}^{\infty} J_{l-m}\left(\alpha\right)
J_{l-n}\left(\alpha\right)~.\label{J1}
\end{eqnarray}
In obtaining the last line in Eq.~(\ref{J1}), we have made use of
the property $J_{n-l}(\alpha)=(-1)^{n-l}J_{l-n}(\alpha)$. To
proceed further, we make use of a version of the addition theorems
of the Bessel functions, namely \cite{AS},
\begin{eqnarray}
\sum_{k=-\infty}^{\infty} J_{k+\nu}\left(\alpha_1\right)
J_k\left(\alpha_2\right)=J_\nu\left(\alpha_1-\alpha_2\right)~.
\end{eqnarray}
It is easy to prove that this theorem leads to the following
interesting identity for Bessel functions with integral orders:
\begin{eqnarray}
\sum_{l=-\infty}^{\infty} J_{l-m}\left(\alpha\right)
J_{l-n}\left(\alpha\right)&=&J_{m-n}\left(0\right) \nonumber\\
&=&\delta_{mn}~. \label{J2}
\end{eqnarray}
Using Eq.~(\ref{J2}), Eq.~(\ref{J1}) becomes $g_n$, which is
exactly the left-hand side of Eq.~(\ref{g}).

Let us now apply the transform to Eq.~(\ref{b.c.B-1}).  We get
\begin{eqnarray}
(-1)^n A^\prime_{n}\sin(k_{n}a) =
 \sum_{l}(-1)^l \left(a^\prime_{l}e^{q_{l}a}+b^\prime_{l}e^{-q_{l}a}
 \right)J_{n-l}(\alpha)~.
 \label{b.c.A-1}
\end{eqnarray}
We recognize that this is the same equation, Eq.~(\ref{b.c.1}),
satisfied by $A_n,~a_n$ and $b_n$, with the same
 Floquet energy $\var$ and frequency $\omega$.  This implies that
 the two sets of coefficients are related by
\begin{eqnarray}
A_n=(-1)^n A^\prime_{n}~~,~~ a_n=(-1)^n
a^\prime_{n}~~,~~b_n=(-1)^n b^\prime_{n}~. \label{coeff-1}
\end{eqnarray}
Similarly, by applying the transform to Eq.~(\ref{b.c.1}), we get
Eq.~(\ref{b.c.B-1}) with the same connections, namely,
Eq.~(\ref{coeff-1}), among the coefficients. With this connection
of the coefficients it is seen that Eqs.~(\ref{b.c.1}) and
(\ref{b.c.B-1}) are dual pairs under the $\cal H$ transform.

Now consider the transform of Eq.~(\ref{b.c.B-3}).  With
Eq.~(\ref{coeff-1}), we get
\begin{eqnarray}
\sum_l (-1)^l t^\prime_{l}e^{ik_{l}b}J_{n-l}\left(\alpha\right)&=&
\sum_l (-1)^l \left(
a^\prime_{l}e^{q_{l}b}+b^\prime_{l}e^{-q_{l}b}\right)
J_{n-l}(\alpha)~,\nonumber\\ &=& \sum_l \left(
a_{l}e^{q_{l}b}+b_{l}e^{-q_{l}b}\right) J_{n-l}(\alpha)~.
\end{eqnarray}
This is the same as Eq.~(\ref{b.c.3}) with the same Floquet energy
and frequency, provided that the coefficients $t_n$ and
$t^\prime_n$ are related by the $\cal H$-transform in the
following forms:
\begin{eqnarray}
t_n e^{ik_n b}&=&\sum_{l=-\infty}^{\infty} (-1)^l t^\prime_l
e^{ik_{l}b} J_{n-l}\left(\alpha\right)~,\nonumber\\ t^\prime_n
e^{ik_{n}b} &=&\sum_{l=-\infty}^{\infty} (-1)^l t_l e^{ik_{l}b}
J_{n-l}\left(\alpha\right)~.
 \label{coeff-2}
\end{eqnarray}

The above discussions show that,  the two models are mapped into
one another under the $\cal H$-transform, with the relevant
coefficients in the wave functions being related through
Eqs.~(\ref{coeff-1}) and (\ref{coeff-2}). Hence the Floquet energy
spectra as function of oscillation's amplitude and frequency of
the two models are identical.  The wave functions, however, are
different, since they have different forms, namely,
Eqs.~(\ref{wf-2})-(\ref{wf-3}) for Model A, and
Eqs.~(\ref{wf-B1})-(\ref{wf-B3}) for Model B.  Nevertheless, it
turns out that the relations between their coefficients in the
region $0\leq x \leq b$, where the particle is confined, can be
understood as the result of a gauge transformation of the two wave
functions. This will be discussed in the next section.

\section{Gauge equivalence of the two models}

In this section we show the underlying connection between the two
models based on gauge invariance principle.

One needs only to consider the region in $0<x\leq b$ , since the
metastable system in these two models is mainly confined within
this  region. The region $x>b$ does not affect the essential
physics underlying the two models. This is because for a
metastable state, its wave function is extremely small (nearly
identical to zero, as confirmed numerically) in the region $x>b$
before the system has actually decayed. This region serves only as
an ideal drain.

The gauge transformation of a quantum system with an
$x$-independent gauge function $\chi (t)$ is given by:
\begin{eqnarray}
\psi(x,t)&\to& \psi^\prime =
e^{\frac{i}{\hbar}\chi(t)}\psi(x,t)~,\nonumber\\ V(x,t)&\to&
V^\prime(x,t) = V-\frac{\partial}{\partial t} \chi(t)~.
\label{gauge}
\end{eqnarray}
When the gauge function $\chi(t)$ is chosen to be
\begin{eqnarray}
\chi(t)=\frac{V_1}{\omega}\sin(\omega t)~, \label{chi}
\end{eqnarray}
the potential in region $0\leq x\leq b$ in Model A,
Eq.~(\ref{potential}), is transformed into
\begin{eqnarray}
V^\prime(x,t)=\left\{
  \begin{array}{ll}
    -V_{1}\cos(\omega t)~, & 0\leq x<a~, \\
    V_{0}~, & a\leq x\leq b~.
  \end{array} \right.
\end{eqnarray}
This potential is physically equivalent to the potential
describing Model B in the same region: they only differ in the
origin of the time variable $t$ by an amount $\pi /\omega$. Hence,
the potentials of the two models are related by a gauge
transformation with gauge function Eq.~(\ref{chi}) and a time
shift $t\to t+ \pi/\omega$.

We now examine the connection between the wave functions of the
two models in the regions considered. Under the gauge
transformation described by Eqs.~(\ref{gauge}) and (\ref{chi}),
the wave function of Model A transforms according to
\begin{eqnarray}
\Psi(x,t)\to \Psi^\prime = e^{iV_1\sin(\omega
t)/\hbar\omega}\Psi(x,t)~,
 \label{wf1-g}
\end{eqnarray}
with
\begin{eqnarray}
 e^{iV_{1}\sin(\omega t)/\hbar\omega}&=&\sum_{m=-\infty}^{\infty} (-1)^m J_m
 \left(\alpha\right)
 e^{-im\omega t}
 \end{eqnarray}
which can be easily obtained from Eq.~(\ref{Bessel}) by changing
$V_1$ to $-V_1$ and using the fact $J_n(-\alpha)=(-1)^n J_n
(\alpha)$. According to the previous discussion, we shall make the
shift $t\to t+ \pi/\omega$ in $\Psi^\prime$ and compare the
resulted wave function $\Psi^\prime (x, t+\pi/\omega)$ with that
of Model B.

First we consider the region $0\leq x\leq a$.  The wave function
of Model A in this region, Eq~(\ref{wf-1}), is gauge-transformed
into
\begin{eqnarray}
 \Psi_{I}^\prime(x,t) & =&e^{iV_1\sin(\omega
t)/\hbar\omega} \sum_{n}A_{n}\sin(k_{n}x)e^{-iE_{n}t/\hbar}
\nonumber\\
 &=& \sum_m \sum_{n}(-1)^m J_m \left(\alpha\right)
A_{n}\sin(k_{n}x)e^{-i(\var+ (n+ m)\hbar\omega)t}\nonumber\\
 &=& \sum_n \sum_l(-1)^{l-n} J_{l-n}\left(\alpha\right)
A_{n}\sin(k_{n}x)e^{-iE_l t/\hbar}~.
 \label{wf-g1}
\end{eqnarray}
In the last line of Eq.~(\ref{wf-g1}) the index $m$ is changed to
$m=l-n$, and $E_l=\var + (m+n)\hbar\omega$.  With the shift $t\to
t+\pi/\omega$, the time-dependent factor in Eq.~(\ref{wf-g1})
picks up a factor $\exp(-iE_l t/\hbar)=(-1)^l \exp(-i\pi\var
/\hbar\omega)$.  Accordingly the wave function is shifted to
\begin{eqnarray}
\Psi_I^{\prime\prime} (x,t)&\equiv& \Psi^\prime (x,
t+\pi/\omega)~\nonumber\\
 &=& e^{-i\pi\var /\hbar\omega}\sum_l \sum_n (-1)^n
A_{n}\sin(k_{n}x)J_{l-n}\left(\alpha\right)e^{-iE_l t/\hbar}~.
\label{wf-s1}
\end{eqnarray}
It is now clear that the wave function $\Psi_I^{\prime\prime}$ in
Eq.~(\ref{wf-s1}) is equivalent, up to an irrelevant phase factor,
to $\Psi_I^\prime$ of Model B, Eq.~(\ref{wf-B1}), with the
identification of $A_n^\prime =(-1)^n A_n$ as discovered in the
last section.

Now we turn to the wave functions in region $a\leq x\leq b$. The
relevant wave function of Model A, Eq.~(\ref{wf-2}),  is
gauge-transformed to
\begin{eqnarray}
 \Psi^\prime_{I\!I}(x,t) &=&  \sum_m \sum_{n}\sum_{l} (-1)^m \left(a_{l}e^{q_{l}x}+b_{l}
  e^{-q_{l}x}\right)  J_{n-l}\left(\alpha\right)J_m\left(\alpha\right)
  e^{-i(E_n + m\hbar\omega)t/\hbar}~\nonumber\\
  &=& \sum_k \sum_{n}\sum_{l}  \left(a_{l}e^{q_{l}x}+b_{l}
  e^{-q_{l}x}\right)  J_{n-l}\left(\alpha\right)J_{n-k}\left(\alpha\right)
  e^{-iE_k t/\hbar}~,
 \label{wf-g2}
\end{eqnarray}
where we have used $J_{-m}\left(\alpha\right)=(-1)^m
J_m\left(\alpha\right)$, and relabeled the index in the last
expression. By making the same shift in the time variable, the
wave function changes to
\begin{eqnarray}
\Psi^{\prime\prime}_{II} (x,t)&\equiv&
\Psi^\prime_{I\!I}(x,t+\pi/\omega)~\nonumber\\
  &=& e^{-i\pi\var /\hbar\omega}\sum_k \sum_{l}(-1)^k  \left(a_{l}e^{q_{l}x}+b_{l}
  e^{-q_{l}x}\right) e^{-iE_k t/\hbar}
  \sum_{n}J_{n-l}\left(\alpha\right)J_{n-k}\left(\alpha\right)~.
\end{eqnarray}
The sum over $n$ in the last factor is simply the Kronecker delta
symbol $\delta_{lk}$ according to Eq.~(\ref{J2}).  Hence
\begin{eqnarray}
\Psi^{\prime\prime}_{II} (x,t)=e^{-i\pi\var /\hbar\omega}\sum_{l}
(-1)^l \left(a_{l}e^{q_{l}x}+b_{l}
  e^{-q_{l}x}\right) e^{-iE_l t/\hbar}~.
 \label{wf-s2}
\end{eqnarray}
Again, up to the irrelevant phase factor, the wave function
$\Psi^{\prime\prime}_{II}$ is equivalent to $\Psi^\prime_{II}$ of
Model B, with $a_l^\prime=(-1)^l a_l$ and $b_l^\prime=(-1)^l b_l$
as before.

The gauge equivalence of the wave functions of the two Models in
the physically relevant regions provides a physical understanding
of the identity of their Floquet spectra.

Finally, we note here that from such gauge invariance argument it
also follows that the static problem is equivalent to the problem
with the bottom of the well and the top of the barrier oscillating
in phase with the same frequency and amplitude.

\section{Summary}

In this paper we investigate the stability of a particle trapped
in between an infinite wall and square barrier, with either a
time-periodically oscillating barrier or bottom of the well. In
these models we do not restrict the amplitude of the
time-periodically oscillating potential to be small, so a
nonperturbative approach based on the Floquet theory is adopted to
tackle the problem. For each case we derive an equation to
determine the Floquet energy of the system. The imaginary part of
the Floquet energy can be used as a measure of the stability of
the system. It is found that in general the oscillating field
causes a quantum metastable state to decay faster.  But when the
Floquet energies (real part) of two states make a direct crossing,
the more stable of the two states will undergo a
resonance-enhanced decay, whereas the less stable state is
slightly stabilized near the resonance frequency. Beyond the
resonance frequency, the states resume stability of the original
order of magnitude. However, if the crossing is an avoided one,
the two states will switch stability: the less (more) stable state
becomes more (less) stable.  This gives the possibility to
stabilize a metastable state, as discussed in \cite{Ho2004}.  Here
we further found that when the amplitude of the oscillating
potential is increased, a direct crossing could change into an
avoided one.  Thus, our results show that one can manipulate the
stability of different states in a quantum potential by a
combination of adiabatic changes of the frequency and the
amplitude of the oscillating barrier.  The two models we
considered here were found to have identical Floquet energy
spectrum for the same frequency and amplitude of the oscillating
field.  This was explained by a discrete transform which connects
the equations of boundary conditions of the two models, and by a
gauge transformation which maps the wave functions of the two
models. These arguments also guarantee that the model discussed in
\cite{Ho2004} will have the same Floquet spectrum, whether the
oscillating part is the barrier or the bottom of the well.

\bigskip

\begin{acknowledgments}
\vskip 0.5cm

This work was supported in part by the National Science Council of
the Republic of China through Grant No. NSC 92-2112-M-032-015 and
NSC 93-2112-M-032-009.  We would like to thank the referee for a
very enlightening comment which inspired us to prove the gauge
equivalence of the systems discussed here.

\end{acknowledgments}

\newpage
\centerline{\bf Figure Captions}
\begin{description}
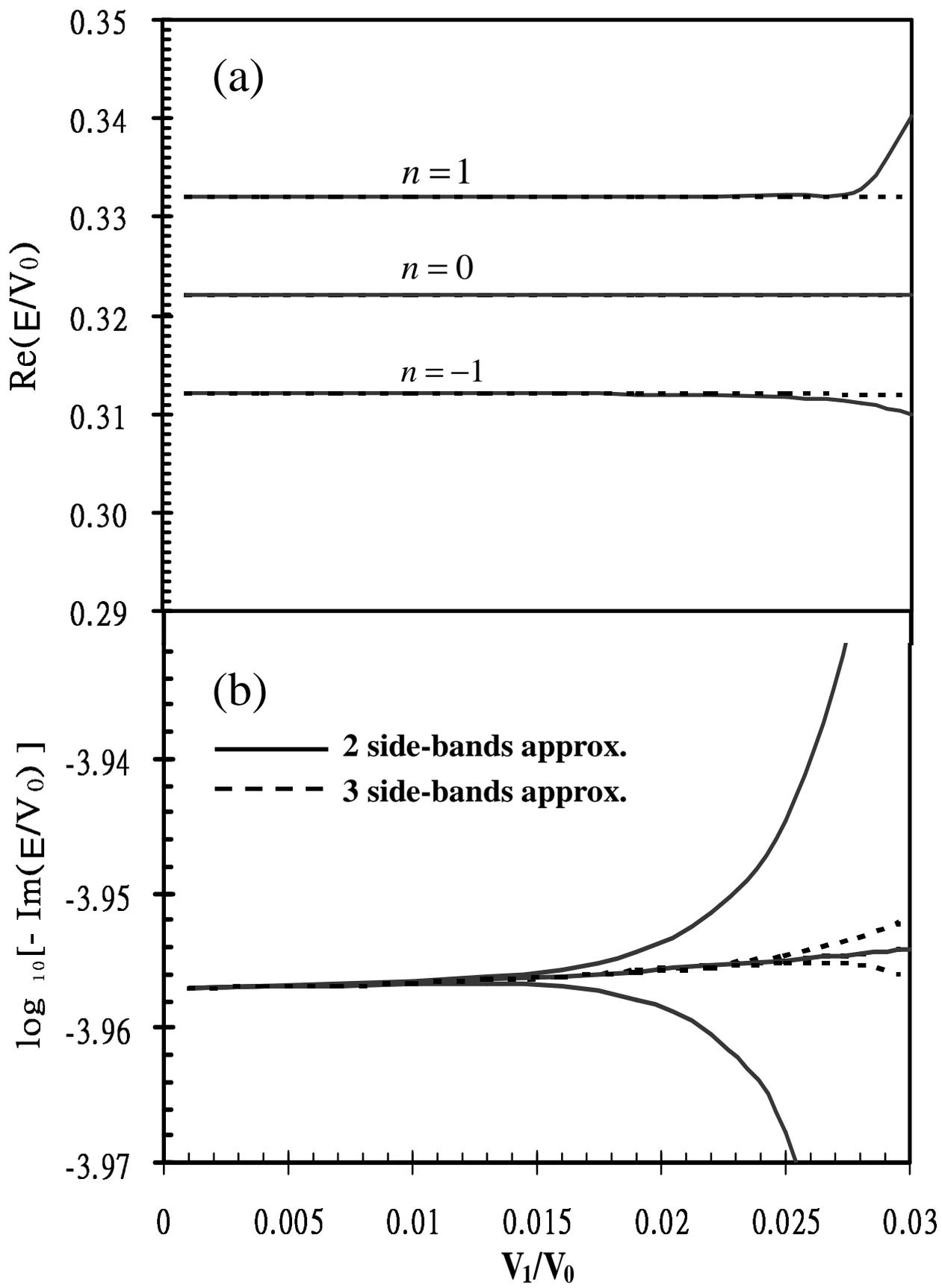
\item[Figure 1.] Relation of the real (a) and imaginary (b)
parts of the Floquet energy vs. the parameter $V_{1}/V_{0}$ at
fixed frequency parameter $\omega/V_{0}=0.01$ for $V_{0}=10.0,
a=1.0, b=2.0 ~(a.u.)$.  Solid and dotted lines  represent the
values evaluated in the 2 side-bands and 3 side-bands
approximation, respectively. Values of $V_1/V_0$ in (b) beyond
which the lines diverge indicate the points of breakdown of the
respective approximations.

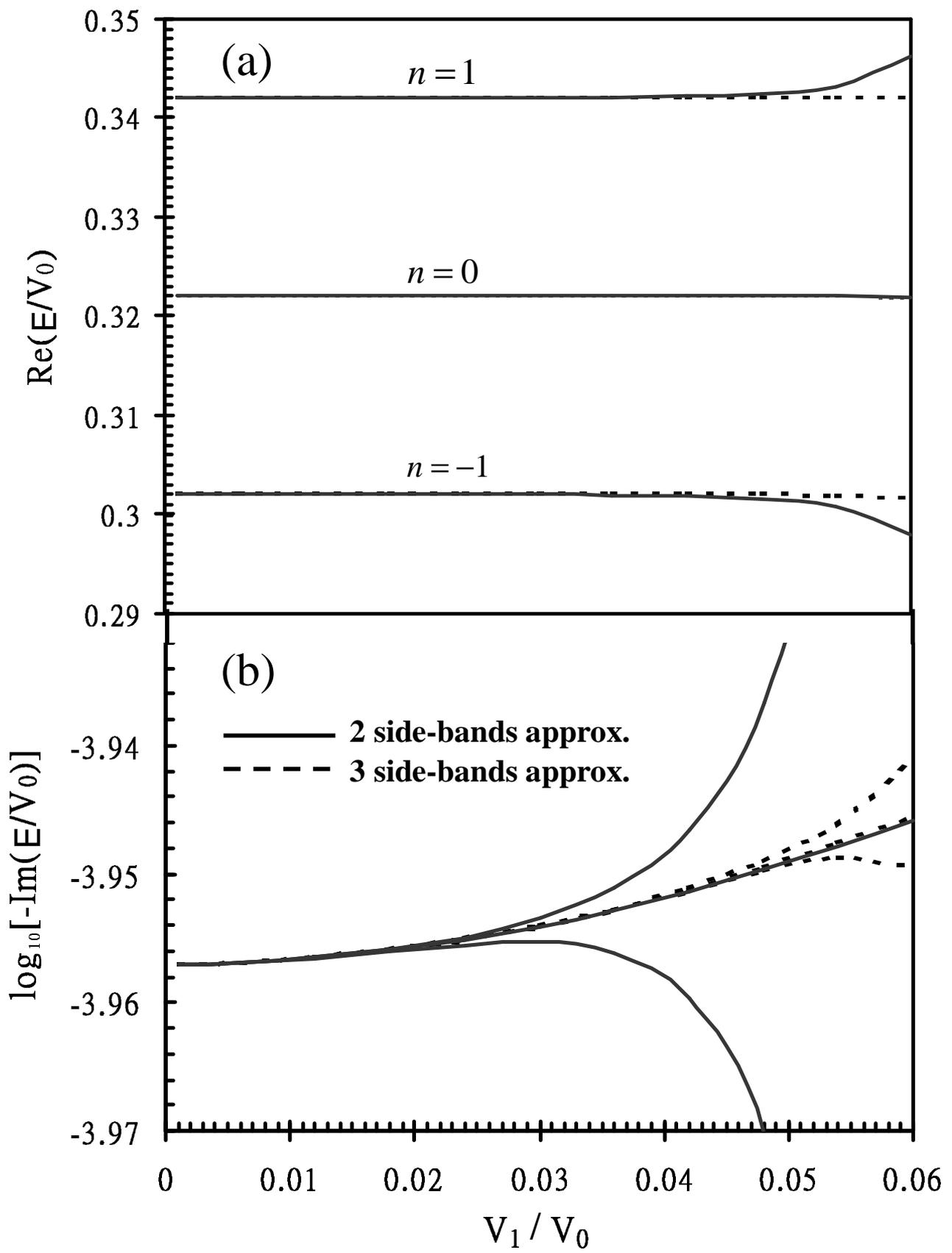
\item[Figure 2.] Same plots as Fig.~1 for $V_{0}=10.0, a=1.0,
b=2.0~(a.u.)$ and the fixed frequency parameter
$\omega/V_{0}=0.02$.

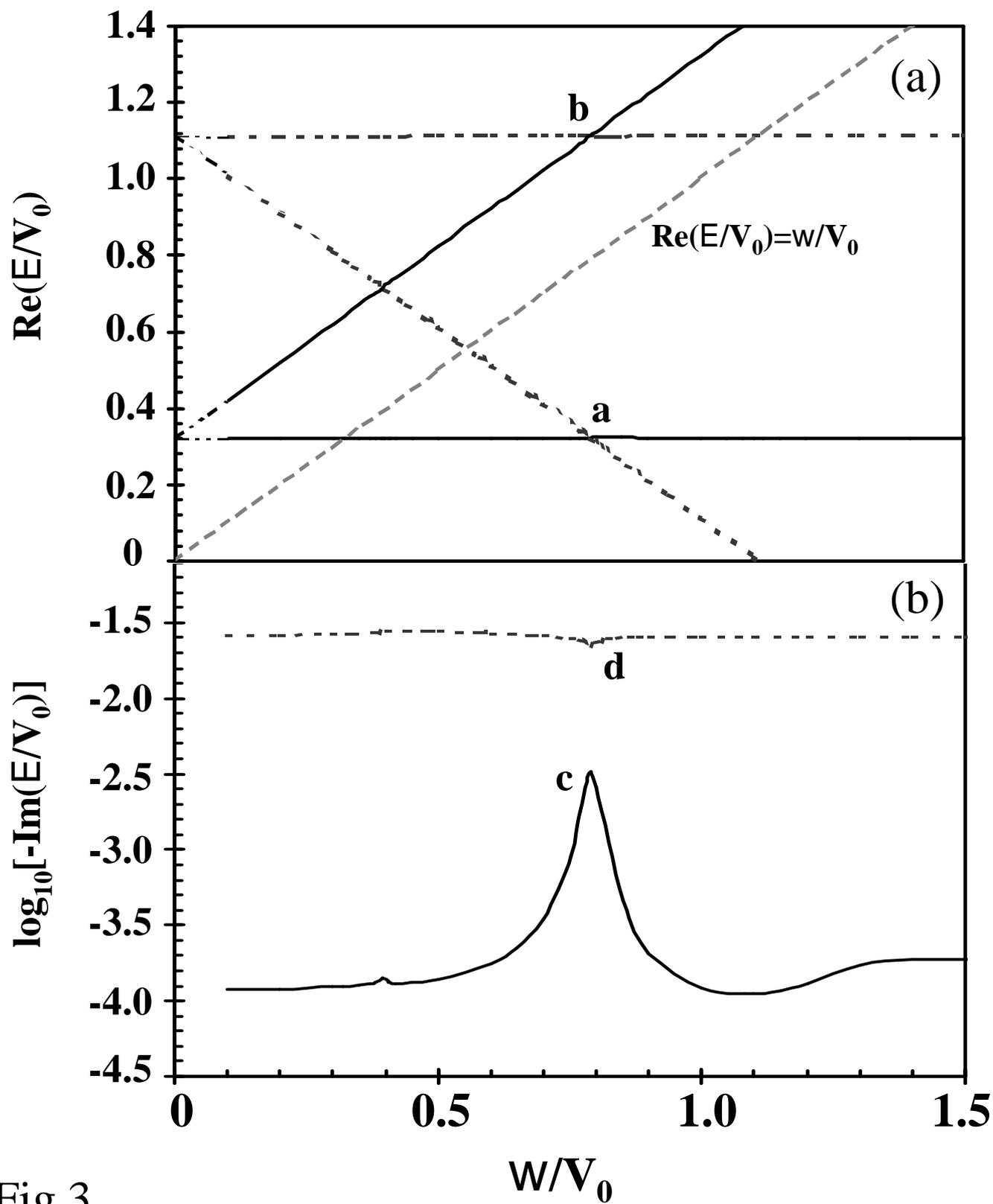
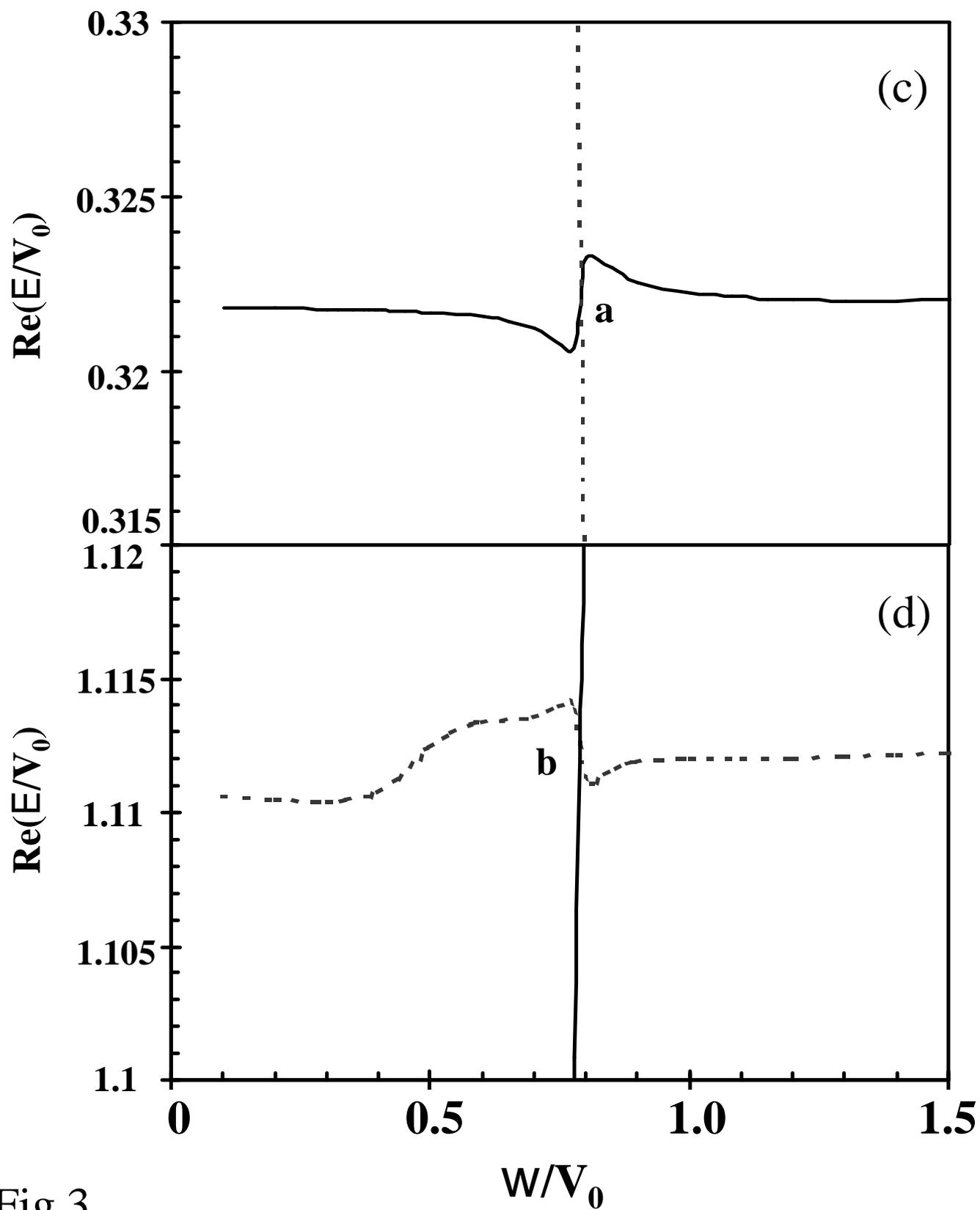
\item[Figure 3.] Relation of the real and imaginary parts of the
 Floquet energy  vs. the the driving frequency $\omega/V_{0}$ for
$V_{0}=10.0$, $V_{1}=1.0$, $a=1.0$ and $b=2.0~(a.u.)$. Fig.~(a)
and (b) show the real and imaginary parts, respectively, of the
Floquet energy. Fig.~(c) and (d)  depict the real part of the
Floquet energy of the lower and the upper state, respectively, at
the direct crossing with a refined scale.

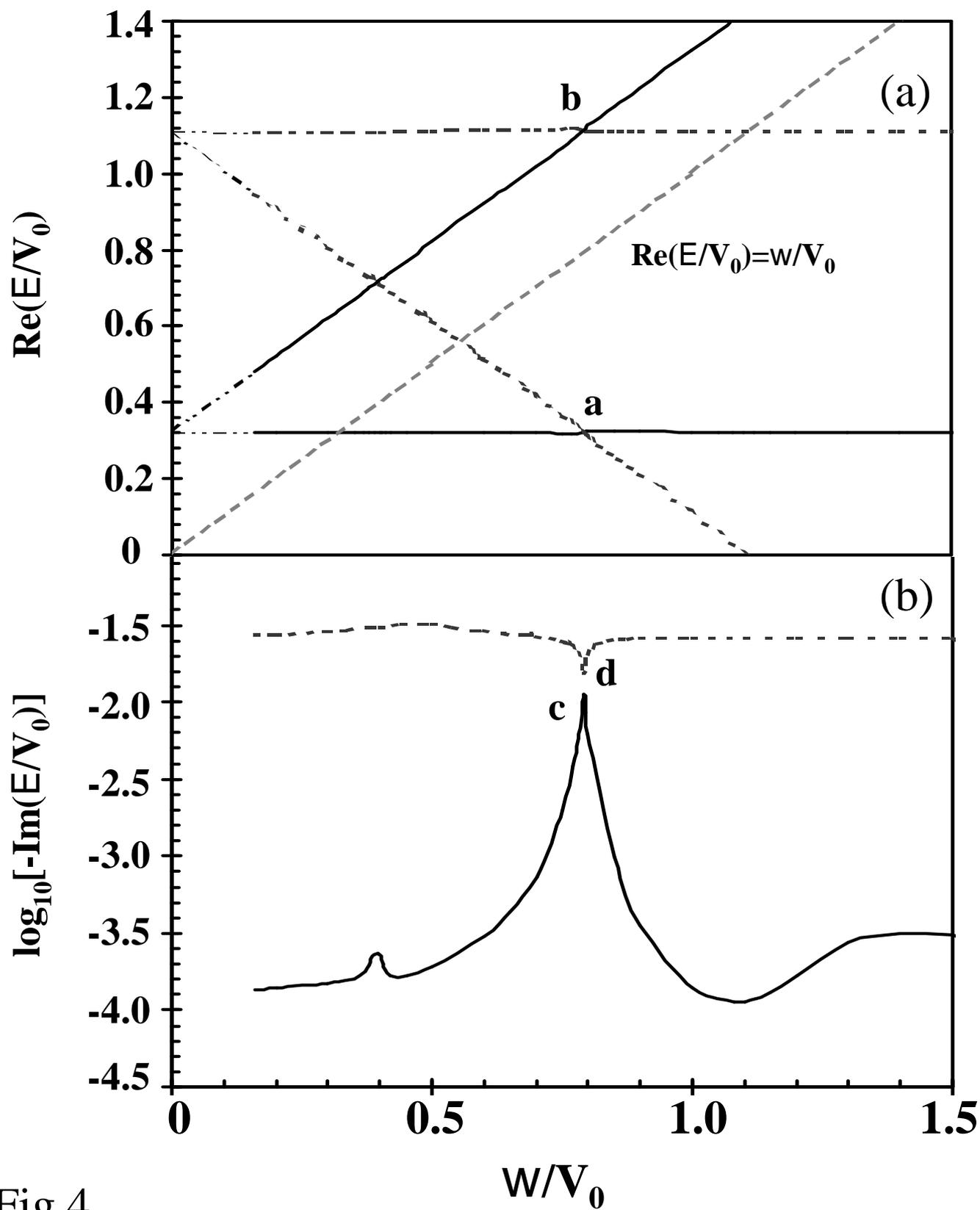
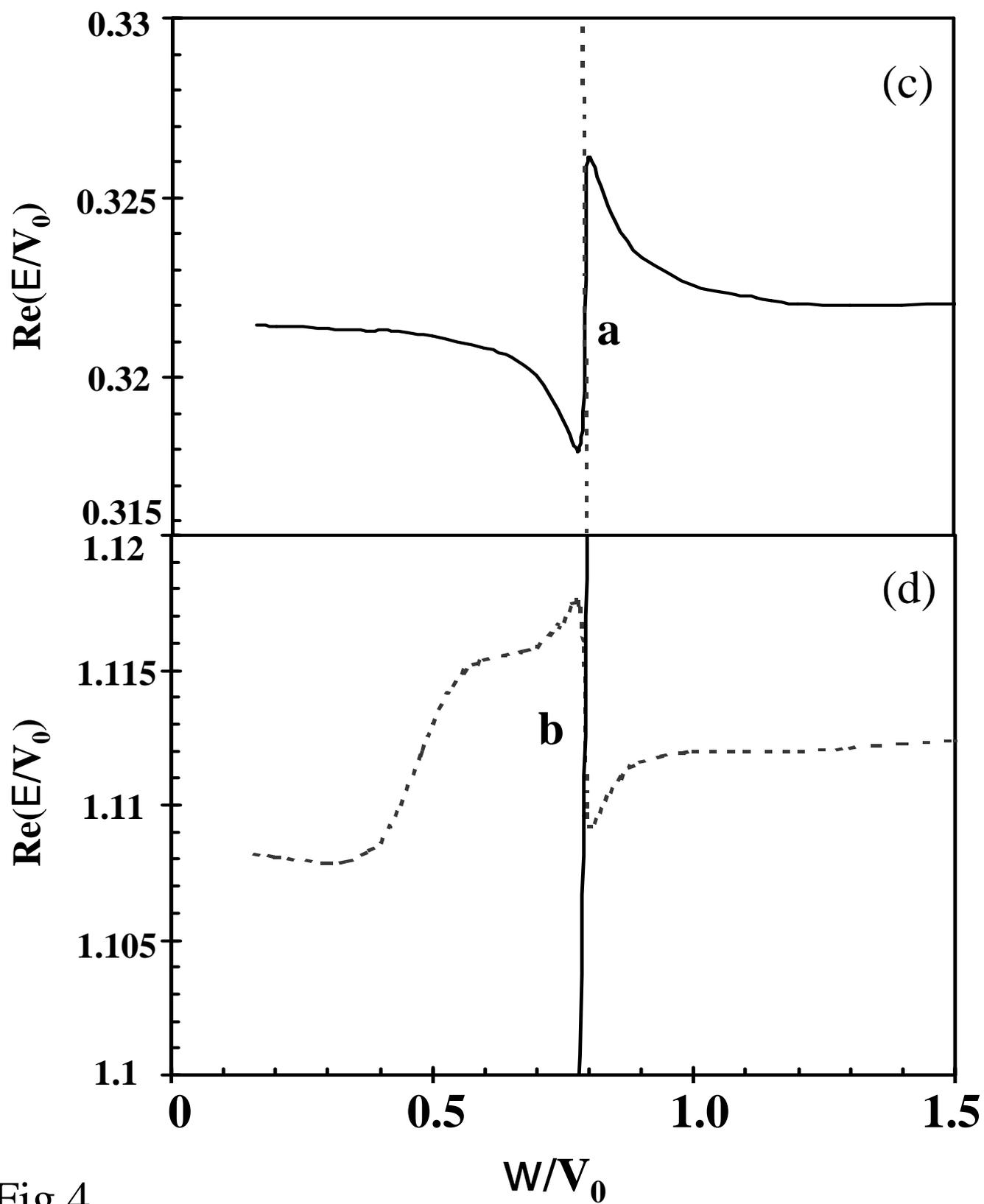
\item[Figure 4.]  Same plots as Fig.~3, but with $V_1=1.6~(a.u.)$.

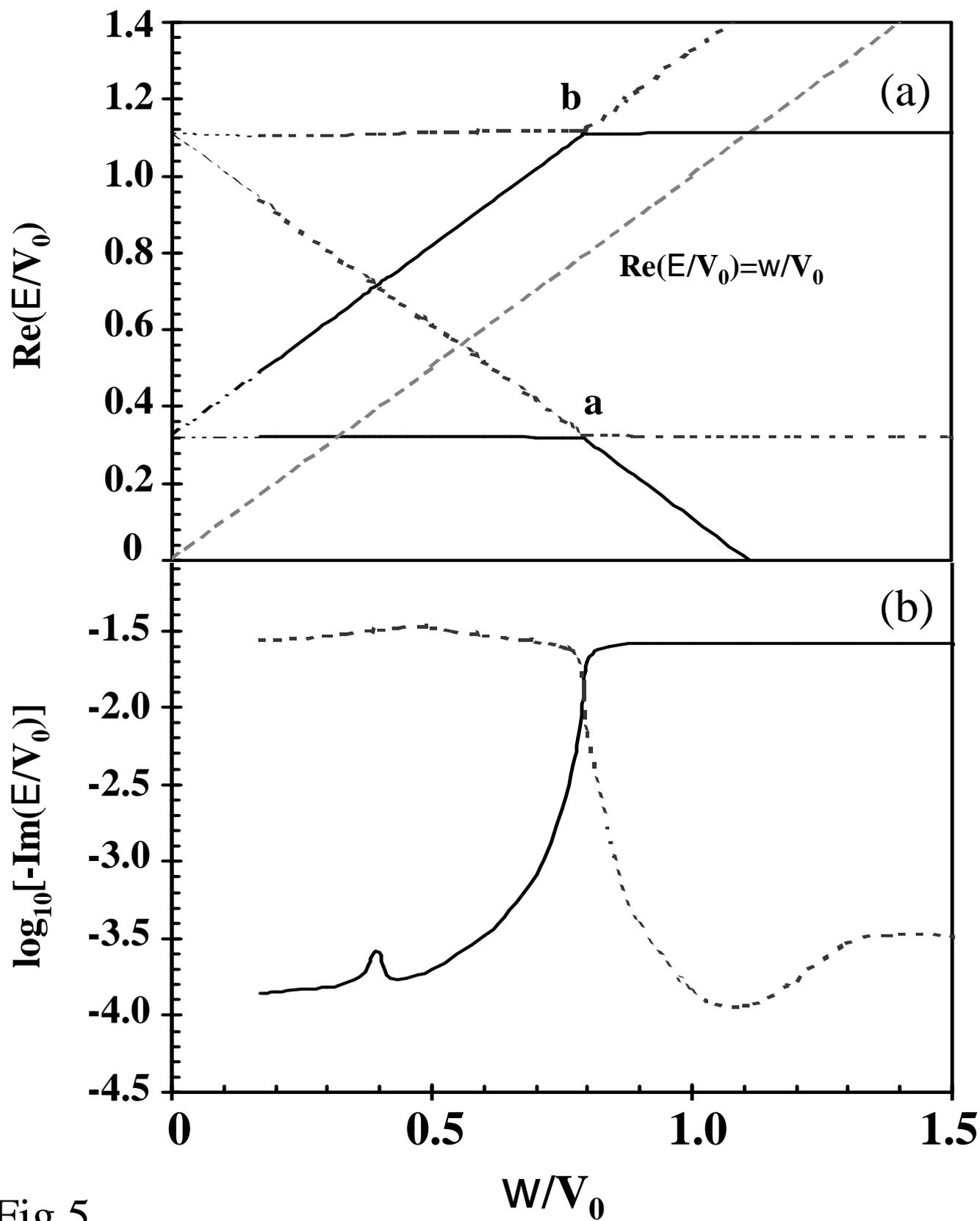
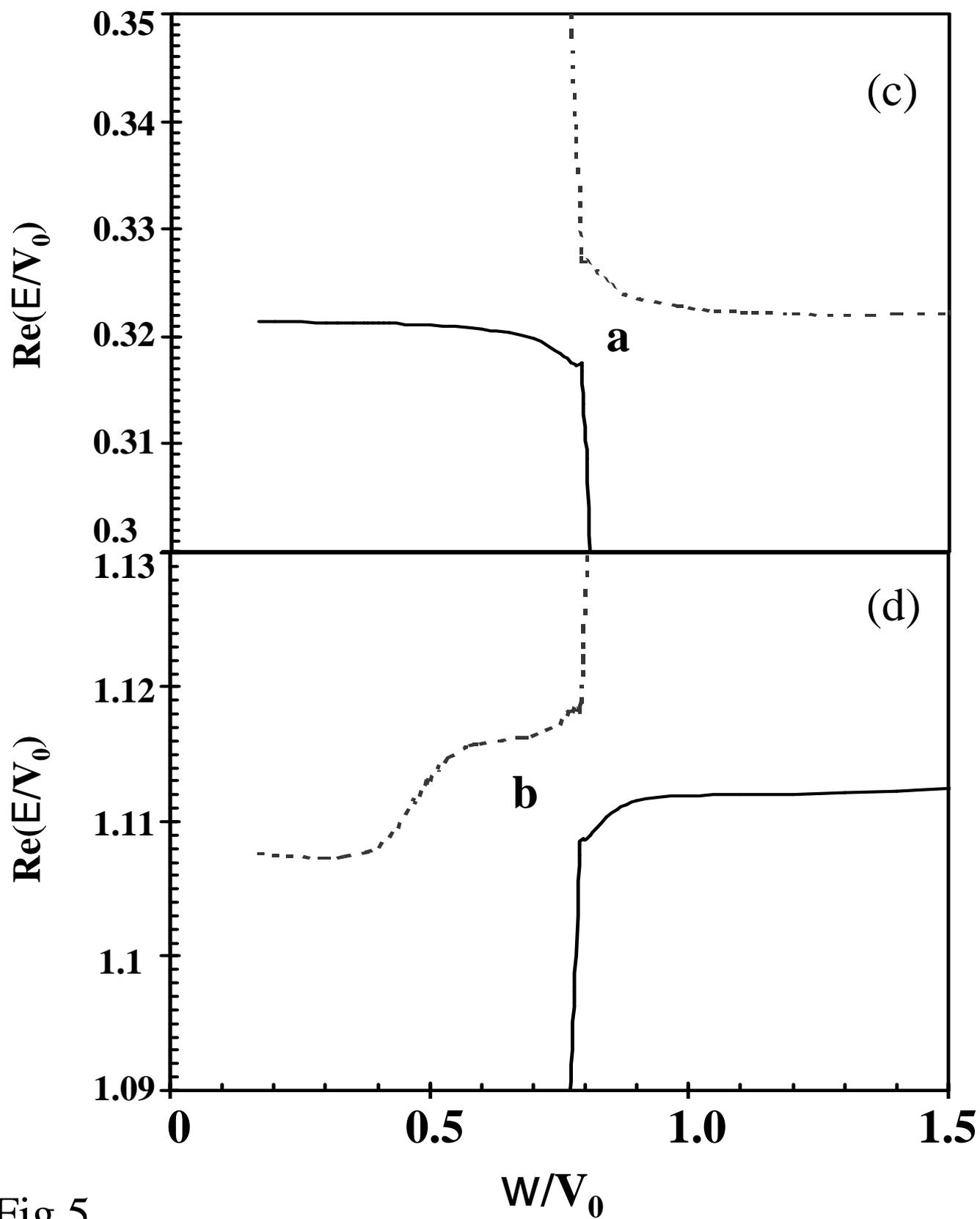
\item[Figure 5.]  Same plots as Fig.~3 with $V_1=1.7~(a.u.)$.  Note
that the crossing has become an avoided one.

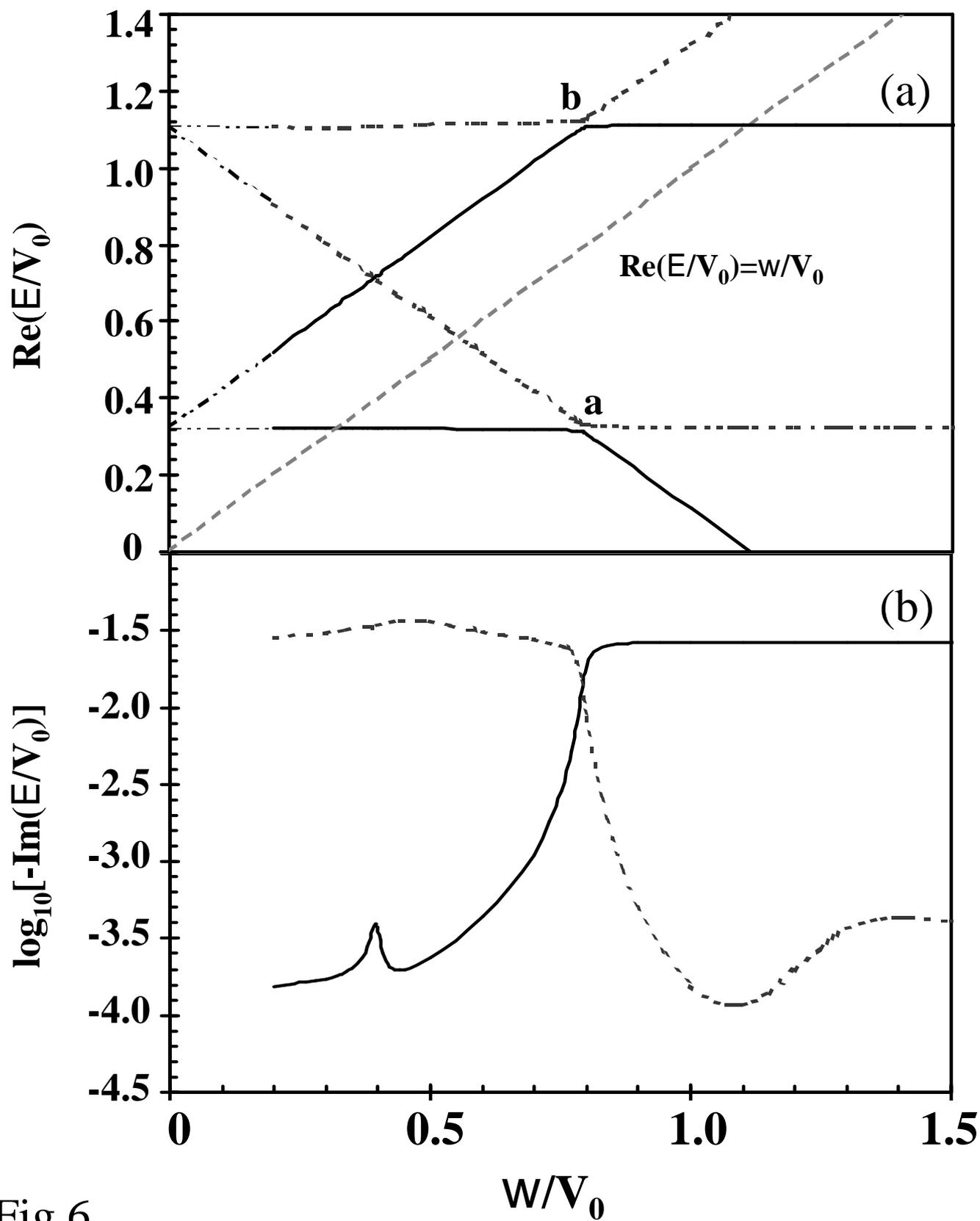
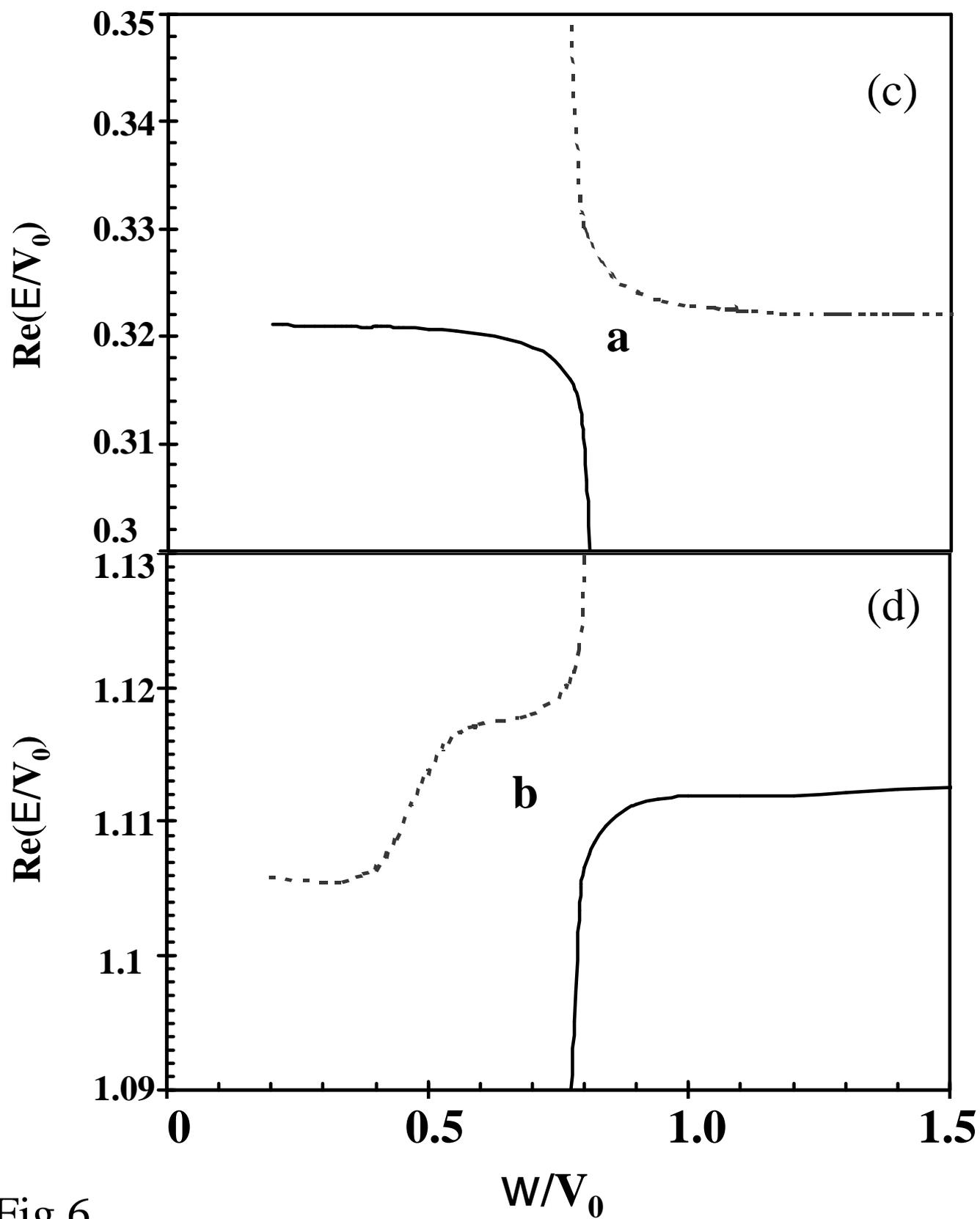
\item[Figure 6.]  Same plots as Fig.~5 with $V_1=2.0~(a.u.)$.

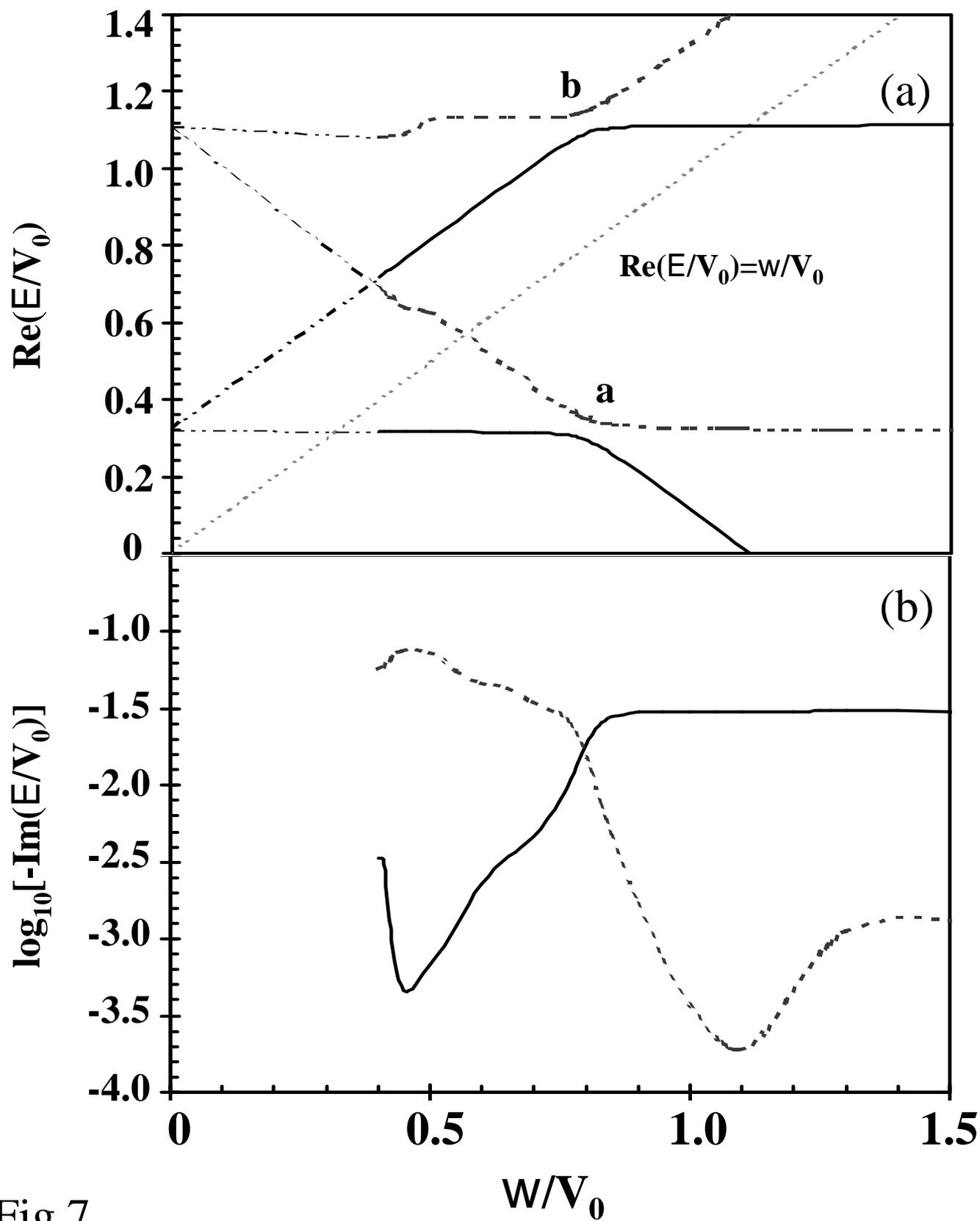
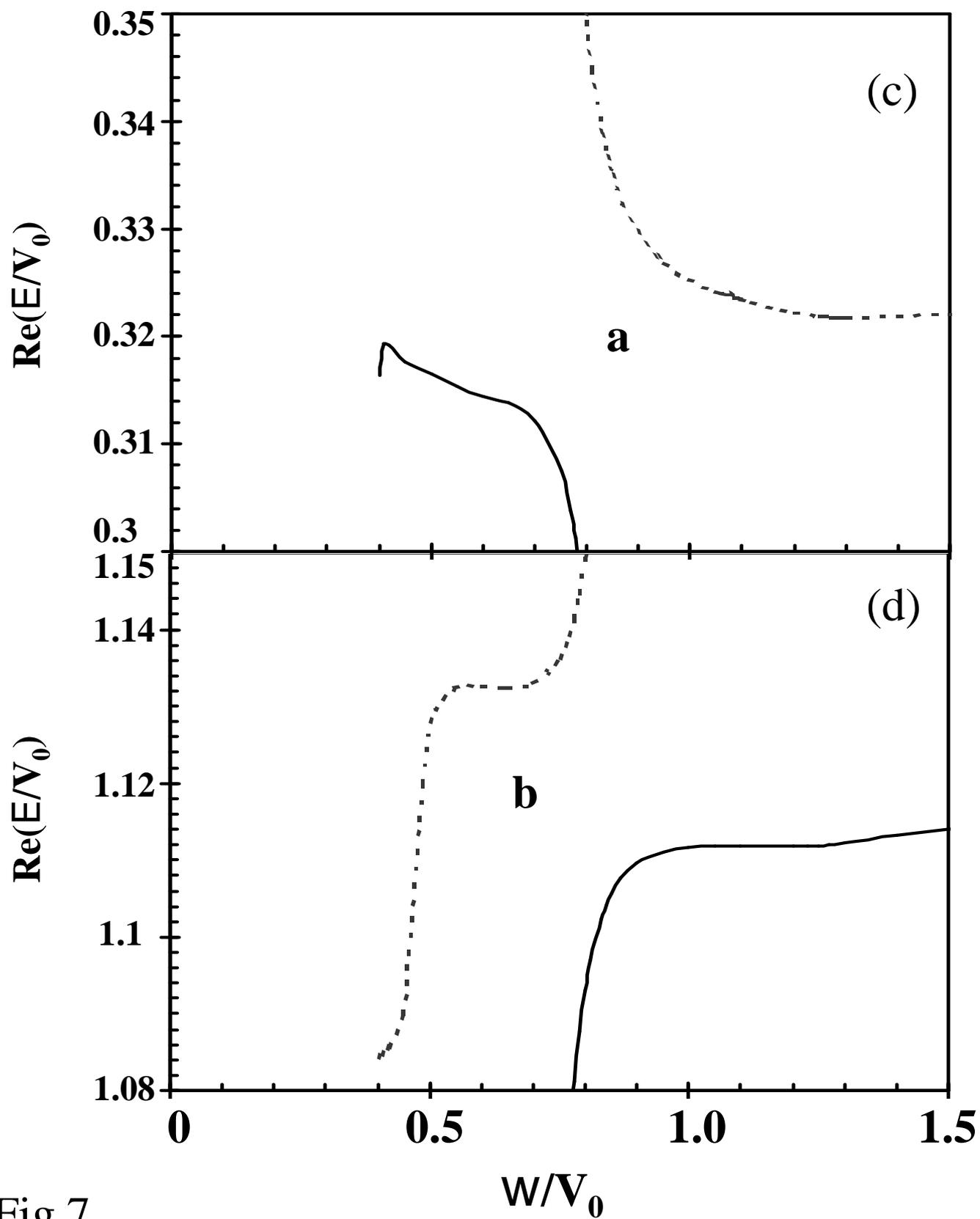
\item[Figure 7.]  Same plots as Fig.~3, but with $V_1=4.0~(a.u.)$.
The avoided crossing is enhanced.

\end{description}

\end{document}